\documentclass[a4paper,onecolumn,accepted=2025-07-21]{quantumarticle}
\pdfoutput=1

\usepackage[utf8]{inputenc}
\usepackage[toc,page]{appendix}
\usepackage{amsfonts}
\usepackage[dvips]{graphicx}
\usepackage{amsmath}
\usepackage{amssymb}
\usepackage{hyperref}
\usepackage{color}
\usepackage{mathrsfs}
\usepackage{isomath}
\usepackage{amsthm}
\usepackage{epstopdf}
\usepackage{txfonts}
\usepackage{dsfont}
\usepackage{ulem}
\usepackage{physics}
\usepackage{tikz} 
\allowdisplaybreaks[4]

\usepackage{bbold}
\usepackage{mathtools}

\newcommand{\be}{\begin{equation}}
\newcommand{\ee}{\end{equation}}
\newcommand{\eea}{\end{eqnarray}}

\renewcommand{\va}[1]{\ensuremath{(\Delta#1)^2}}

\newcommand{\ex}[1]{\ensuremath{\left\langle{#1}\right\rangle}}
\newcommand{\exs}[1]{\ensuremath{\langle{#1}\rangle}}

\newcommand{\KetBra}[1]{\ensuremath{| #1 \rangle \langle #1 |}}
\newcommand{\Ket}[1]{\ensuremath{|#1\rangle}}

\newcommand{\kommentar}[1]{}
\renewcommand{\trace}{{\rm Tr}}

\newcommand{\aver}[1]{\langle #1 \rangle}
\newcommand{\diag}{\mathrm{diag}}

\newcommand{\id}{\openone}
\newcommand{\idmap}{{\rm id}} 

\renewcommand{\vec}[1]{\boldsymbol{#1}}

\newcommand{\rhoavtwo}{\varrho_{\rm av2}}

\newcommand{\btext}[1]{#1}

\newcommand{\Wspace}{\mathcal W}

\newcommand{\OmegaSigma}{\Omega}

\newcommand{\LambdaSigma}{\Lambda}

\newcommand{\barCkl}{(C_{\rm av2})_{kl}}

\newcommand{\barC}{C_{\rm av2}}
\newcommand{\barCvr}{C_{\rm av2}}

\newcommand{\barXkl}{({\gamma}_{\rm av2})_{kl}}

\newcommand{\barX}{\gamma_{\rm av2}}

\newtheorem{observation}{Observation}

\newtheorem{lemma}{Lemma}

\newtheorem{definition}{Definition}

\usepackage{cleveref}
\crefname{equation}{Eq.}{Eqs.}
\crefname{figure}{Fig.}{Figs.}
\crefname{observation}{Observation}{Observations}
\crefname{corollary}{Corollary}{Corollaries}
\crefname{lemma}{Lemma}{Lemmata}
\crefname{proof}{Proof}{Proofs}
\creflabelformat{proof}{#2proof#3}
\crefname{remark}{Remark}{Remarks}
\crefname{prop}{Proposition}{Propositions}
\crefname{definition}{Definition}{Definitions}
\crefname{appendix}{Appendix}{Appendix}

\begin{document}
\title{$su(d)$-squeezing and many-body entanglement geometry in finite-dimensional systems}

\author{Giuseppe Vitagliano}
\orcid{0000-0002-5563-3222}
 \email{giuseppe.vitagliano@tuwien.ac.at}
\affiliation{Vienna Center for Quantum Science and Technology, Atominstitut, TU Wien,  A-1020 Vienna, Austria}

\author{Otfried G\"uhne}
\orcid{0000-0002-6033-0867}
\email{otfried.guehne@uni-siegen.de}
\affiliation{Naturwissenschaftlich-Technische Fakult{\"a}t, Universit{\"a}t Siegen, Walter-Flex-Stra{\ss}e 3, D-57068 Siegen, Germany}

\author{G\'eza T\'oth}
\orcid{0000-0002-9602-751X}
\email{toth@alumni.nd.edu}
\affiliation{Department of Theoretical Physics, University of the Basque Country UPV/EHU, P.O. Box 644, E-48080 Bilbao, Spain}
\affiliation{EHU Quantum Center, University of the Basque Country UPV/EHU, Barrio Sarriena s/n, E-48940 Leioa, Biscay, Spain}
\affiliation{Donostia International Physics Center DIPC, Paseo Manuel de Lardizabal 4, E-20018 San Sebasti\'an, Spain}
\affiliation{IKERBASQUE, Basque Foundation for Science, E-48009 Bilbao, Spain}
\affiliation{HUN-REN Wigner Research Centre for Physics, P.O. Box 49, H-1525 Budapest, Hungary}

\begin{abstract}
Generalizing the well-known spin-squeezing inequalities, we study the relation between squeezing of collective $N$-particle $su(d)$ operators and many-body entanglement geometry in multi-particle systems. For that aim, we define the set of pseudo-separable states, which are mixtures of products of single-particle states that lie in the $(d^2-1)$-dimensional Bloch sphere but are not necessarily positive semidefinite. We obtain a set of necessary conditions for states of $N$ qudits to be of the above form. Any state that violates these conditions is entangled. We also define a corresponding $su(d)$-squeezing parameter that can be used to detect entanglement in large particle ensembles. Geometrically, this set of conditions defines a convex set of points in the space of first and second moments of the collective $N$-particle $su(d)$ operators. We prove that, in the limit $N\gg 1$, such set is filled by pseudo-separable states, while any state corresponding to a point outside of this set is necessarily entangled. We also study states that are detected by these inequalities: We show that states with a bosonic symmetry are detected if and only if the two-body reduced state violates the positive partial transpose (PPT) criterion. On the other hand, highly mixed states states close to the $su(d)$ singlet are detected which have a separable two-body reduced state and are also PPT with respect to all possible bipartitions. We also provide numerical examples of thermal equilibrium states that are detected by our set of inequalities, comparing the spin-squeezing inequalities with the $su(3)$-squeezing inequalities. 
\end{abstract}

\maketitle

\section{Introduction}

Despite decades of efforts, a central problem in quantum information remains how to distinguish a separable state from an entangled one, 
the latter being central for the study of complex many-body systems and also for applications \cite{horodeckirev,amico08,Guhne2009Entanglement,Laflorencie16,FriisNatPhys19}. The task becomes extremely difficult especially when the number of particles grows toward macroscopic ensembles, that is the paradigmatic situation with gases and condensed matter systems and
also the target of many current experiments~\cite{pezzerev18}. 
A possible approach to deal with very large number of particles, e.~g., $10^3-10^{12}$ or larger, is to address collectively the ensemble by measuring level populations~\cite{FriisNatPhys19,FrerotFadelLewenstein23}. From these measurements one can directly extract information about interparticle correlations, thus potentially providing information about interparticle entanglement. 

Further simplifications usually come from exploiting the symmetry of the system. In particular, to detect very efficiently multiparticle spin states with permutation symmetry, so-called spin-squeezing criteria for entanglement have been derived in connection with uncertainty relations of spin components, based on an analogy with bosonic quadratures~\cite{Ma2011Quantum}. Most famously, a seminal entanglement criterion valid for systems of $N$ spin-$1/2$ particles has been derived in terms of a spin-squeezing parameter~\cite{Kitagawa1993Squeezed,Wineland1994Squeezed,Sorensen2001Many-particle,Sorensen2001Entanglement}, 
which also directly points at entanglement as a resource for quantum metrology: spin-squeezing signals both interparticle entanglement and a sensitivity of estimating phases coming from rotations enhanced over separable states~\cite{pezzerev18,tothapellaniz14}. 
The original definition of spin squeezing, requiring a large spin polarization, has been generalized over the years in multiple ways, especially towards exploring further connections between symmetries of the target state, entanglement and quantum metrology~\cite{Ma2011Quantum,pezzerev18,tothapellaniz14}. 
In particular, singlet states~\cite{tothmitchell10,behbood14,Kong2020Measurement-induced,urizar13}, Dicke states~\cite{Wieczorek2009Experimental,Prevedel2009Experimental,Lucke2011Twin,Hamley2012Spin-nematic,Lucke2014Detecting,vitagliano16,Luo620,Hoang2016Characterizing} and planar squeezed states \cite{He2011Planar,1367-2630-14-9-093012,puentes13,ColangeloPQS2017,PhysRevA.97.020301} have been produced experimentally and studied with spin-squeezing methods. Note that, unlike the original spin squeezed states, Dicke states and singlet states do not have a large spin polarization.
More recently, experiments have also been performed in which spin-squeezed states are split in spatially separated parts and entanglement has been detected between the parts~\cite{Fadel18,KunkelEtAl2018,Lange18}. Moreover, spin-squeezing has been also connected with entanglement monotones~\cite{FadelVitagliano_2021,liu2022bounding}, and very recently, it has been defined also in the context of randomized measurements \cite{Imai2023Collective}.

Purely from the point of view of entanglement theory, spin-squeezing criteria are related to the criteria based on two-body correlations, especially those coming from local uncertainty relations \cite{hofman03,PhysRevA.74.010301,PhysRevA.81.012324,2017arXiv170510679S,giovannettietalPRA03}, like the covariance matrix criterion \cite{guhnecova,gittsovich08,PhysRevA.82.032306}. 
In particular, a closed set of inequalities, linear in the variances and second moments of the collective spin components has been derived \cite{tothPRL07,tothPRA09}, and generalized for all spin-$j$ particle systems \cite{vitagliano11,vitagliano14}, which highlights the geometry of the set of separable states in the space of $N$-particle spin variances. 
This set of generalized spin-squeezing inequalities was in fact shown to define a convex polytope in the space of $N$-particle collective second moments, for a given value for the first moments, that for $N\gg 1$ is completely filled by separable states \cite{tothPRA09,vitagliano14}. Thus, for large $N$ no more conditions based on the variances and expectation values of collective $N$-particle spin components can be found that can detect a wider set of entangled states.

How could we generalize further the spin-squeezing approach by exploring further the geometry of separable states in the space of correlations accessible by collective measurements? If we want to extend the completeness property mentioned above to all possible collective measurements, then we have to look at quantities other than second moments of the $N$-particle spin components. For systems of particles with more than $d=2$ levels, i.~e., with spin $j=(d-1)/2$ larger than $1/2$, the set of $N$-particle observables is much wider than the three collective spin components  \cite{Duan2000Squeezing,Mustecaplioglu2002SpinSqueezing,Klimov2011Qutrit}, which has been already used, e.g., in cold atom experiments~\cite{Hamley2012Spin-nematic,Stamper_KurnRev13}.
Thus, one way to proceed by keeping the focus on variances of $N$-particle collective quantities, which is also desirable from an experimental perspective and for potential connections with quantum metrology, is to consider observables different from spin components~\cite{vitagliano11}. See also Ref.~\cite{M_ller_Rigat_2022} for a similar approach.
In particular, it makes sense to consider observables belonging to the $su(d)$ algebra, a basis of which is composed by $(d^2 -1)$ traceless Hermitean operators. Thus, for a system of $d$-level particles with $d>2,$ the squeezing of collective $su(d)$ observables is not necessarily equivalent to squeezing of spin observables, i.e., spin squeezing.

In fact, in Ref.~\cite{vitagliano11} a closed set of necessary conditions for separability analogous to the ones in Refs.~\cite{tothPRA09,vitagliano14} was found, and it was shown that interesting
states can be detected, such as many-body $su(d)$ singlet states. However, these relations are valid for each given set of $su(d)$ generators. 
It might happen that a certain quantum state is detected as entangled for one choice of $su(d)$ generators, while it is not detected as entangled for another choice of $su(d)$ generators and, unlike in the $d=2$ case where essentially all single particle bases are equivalent to the three Pauli matrices, it is thus a nontrivial question to
determine the best possible choice of single-particle operators.
Here we address this problem, and at the same time extend the results of Ref.~\cite{vitagliano11} in several directions. First of all, we show that a similar approach can be extended to all basis of $d$-particle observables, not just those composed by $su(d)$ generators. We still call this the set of {\it $su(d)$-squeezing inequalities}, because later we show that the choice of basis where a $su(d)$ basis appears is in fact optimal. 

We do this by deriving an expression that is independent of the single-particle operator basis chosen and allows to obtain the optimal inequality to detect a given state $\varrho$ as entangled.
Thus, on the one hand we solve the problem of determining the optimal set of $su(d)$ generators, if we want to detect a given quantum state as entangled.
This defines an entanglement parameter that we call {\it $su(d)$-squeezing parameter}. 
On the other hand, we represent the inequalities geometrically as a convex set in the space of collective second moments, and the entanglement parameter acquires the interpretation of a signed distance to such a set.  
On top of that, after defining a class of states that we call pseudo-separable, that resembles (and includes) the traditional separable states, we show a completeness property very similar to the spin-squeezing case: we prove that the set of $su(d)$-squeezing inequalities defines a convex polytope that is completely filled by such pseudo-separable states. In particular, its extremal points are associated to pseudo-separable matrices, many of which are not positive, but have a form very similar to those of the extremal states of the spin-squeezing polytope.

Afterwards, we connect our inequalities to the properties of two-body marginal states and show that our parameter can be expressed in terms of correlations extracted from the average two-particle reduced density matrix. 
Finally, we study what states can be detected. We show that permutationally symmetric, i.~e., bosonic states 
are detected if and only if the partially transposed two-particle reduced density matrix has some negative eigenvalues. In other words, 
such states are detected equivalently by our parameter and the so-called Positive Partial Transpose entanglement criterion evaluated on the two-particle reduced state. 
We will formulate this statement more precisely and prove it later in the paper. 
However, we also show that highly mixed states close to the $su(d)$-singlet are detected as entangled which have separable two-body reduced states and are also PPT with respect to all possible bipartitions. 
Finally, we numerically investigate thermal states of $su(3)$ as well as spin-$1$ $N$-particle models for small values of $N$ and compare our $su(3)$-squeezing inequalities with the spin-squeezing inequalities derived in Ref.~\cite{vitagliano14}.

The paper is organized as follows. In \cref{sec:preliminary}, we introduce our framework. In \cref{sec:collcovs},  we define the $N$-particle correlation matrices that will be the central object from which we derive entanglement conditions and discuss how those are related with average two-body correlations. In \cref{sec:polyspinsqueez},  we recall the set of spin-squeezing inequalities derived in \cite{tothPRA09,vitagliano14} and discuss their basic properties, focusing in particular on the geometrical aspects. There, we also introduce other basic bipartite entanglement criteria, namely the so-called Positive Partial Transpose (PPT) \cite{peresPPT,Horodecki19961} and Computable Cross-Norm and Realignment (CCNR) \cite{Rudolph2003,chen03,Rudolph_2005} criteria. Finally, in \cref{sec:nonphysical},  we introduce the definition of pseudo-separable matrices.

Afterwards, we present our results. In \cref{eq:sudcriteriamain},  we derive the set of $su(d)$-squeezing inequalities extending the approach of Ref.~\cite{vitagliano11} to include all possible single-particle operator bases and also become basis-independent. Afterwards, we derive an optimal {\it $su(d)$-squeezing parameter} that could detects a given state $\varrho$ as entangled and is also independent of global changes of $su(d)$ basis. 
In \cref{sec:geometry},  we discuss the geometry of the set of $su(d)$-squeezing inequalities. First we define the variable of our geometric construction in \cref{sec:sudvariablesgeom}, which are given in terms of $N$-particle covariances. Afterwards, we present the representation of our inequalities as a convex set in this space in \cref{sec:polyspinsqueez} and in \cref{sec:completenessmain} we show that our parameter detects all pseudo-separable states that can be detected based on $N$-particle $su(d)$ averages and covariances. Then, in \cref{sec:SSpolytopecomp} we provide more details on the relation between our geometric construction and that of the spin-squeezing inequalities appeared in Ref.~\cite{tothPRA09}.

In \cref{sec:sudsqandtwobody}, we present the expression
of our parameter in terms of average two-particle correlations and discuss what the relation is between our set of inequalities and the entanglement of the two-body reduced state.
In particular, we show that for bosonic states our set of conditions is equivalent to the PPT and CCNR criteria evaluated on the average two-body reduced density matrix, while for states
that are permutationally invariant but not bosonic this is not true. In fact, we show that states in the vicinity of a $su(d)$ singlet state are detected even if they have separable two-body marginals.

In \cref{eq:sudsq}, we investigate numerically what states are detected by our parameter, focusing on thermal states of $su(3)$ models and $su(2)$ models in the spin-$1$ representation with two-body interactions in fully connected graphs which are invariant under permutation of the particles (but not necessarily bosonic). We first observe that states close to the $su(3)$-singlet are indeed detected such that they are PPT with respect to all bipartitions even for small particle numbers. We first observe that states close to the $su(3)$-singlet are indeed detected such that they are PPT with respect to all bipartitions even for small particle numbers. Afterwards, focusing on spin-$1$ models we can find examples of thermal states that are detected by either the set of $su(3)$-squeezing or the spin-$1$-squeezing inequalities of \cite{vitagliano14}, but not by both of them. 
This also implies that in order to fully characterize entanglement in higher spin ensembles, even for just permutationally invariant states, different sets of squeezing inequalities are needed for different sets of operators. It is thus an interesting open question what is the complete set of conditions for, e.g., a spin-$1$ ensemble. See also~\cite{M_ller_Rigat_2022} for another approach that provides further insights.
Finally, in \cref{sec:conclusions}, we conclude summarizing our results and presenting some interesting future directions.

\section{Preliminary tools}\label{sec:preliminary}

In this section, we introduce some preliminary concepts useful for the rest of the discussion.
We define the $N$-particle operators used in our entanglement conditions, as well
as non-physical quantum states used to derive separability bounds and also introduce previous entanglement criteria such as, in particular, the complete set of spin-squeezing inequalities of Refs.~\cite{tothPRA09,vitagliano14}.

\subsection{Variances of $N$-particle observables and average $2$-body correlations for a particle ensemble}\label{sec:collcovs}

Every density matrix of a single $d$-level particle can be expressed as 
\be
\varrho = \frac 1 d \id +  \frac1 2\sum_{k=1}^{d^2-1} \aver{g_k}_\varrho g_k,\label{eq:rhodef}
\ee
where $\aver{\cdot}_\varrho=\trace(\cdot \varrho)$ is the expectation value \footnote{To simplify the notation, we will often omit the subscript $\varrho$ and just write $\aver{\cdot}$ if the state is clear from the context.} and $\{ g_{k} \}_{k=1}^{d^2-1}$ is a set of $su(d)$ generators, i.e, \cite{Vitagliano2011Spin}
\be
\trace(g_k g_l)= 2\delta_{kl}\label{eq:ortog}
\ee
holds and the $g_k$ are Hermitian operators. The advantage of choosing the normalization given in Eq.~\eqref{eq:ortog} is that for $d=2,$ the Pauli spin matrices are such generators. For $d=3,$ the Gell-Mann matrices are also such $su(d)$ generators. The operators $g_k$ together with an operator proportional to the identity can also be called a Local Orthogonal Basis (LOB), however, in that case they typically use a different normalization.

The constraint on the density matrix to be a positive operator translates into a number of conditions on $\aver{\vec g} :=(\aver{g_1},\dots,\aver{g_{d^2-1}})$, the simplest of which is 
\be\label{eq:absvecg}
\aver{\vec g}^2 = \sum_{k=1}^{d^2-1} \aver{g_k}^2 \leq \Lambda_{\max},
\ee
that comes from the upper bound on the purity
\begin{equation}\label{tracerho2}
{\rm Tr}(\varrho^2)\le 1.
\end{equation}
Here, we define the square of the maximal length for $\aver{\vec g}$ as
\be
\Lambda_{\max}=2\frac{d-1}{d}.
\ee
The same constraint can be also interpreted by means of a Local Uncertainty Relation as
\begin{equation}\label{localunc}
(\Delta \vec g)_\varrho^2 := \sum_{k=1}^{d^2-1} (\Delta g_k)_\varrho^2 = \sum_{k=1}^{d^2-1}  \left( \aver{g_k^2}_\varrho - \aver{g_k}^2_\varrho \right) \geq 2(d-1) ,
\end{equation}
simply because the relation
\begin{equation}
\sum_{k=1}^{d^2-1} g_{k}^2 = (d+1) \Lambda_{\max}\cdot\id
\end{equation}
holds for a summation over all the $su(d)$ generators.

Next, we define some observables describing ensembles of $N$ particles that will be used in our paper.
First of all, we define the $N$-particle operators given as a sum of single-particle observables 
\be\label{eq:Sigmak}
G_k=\sum_{n=1}^N g_k^{(n)} ,
\ee
\btext{where $g_k^{(n)}$ is the local $su(d)$ operator $g_k$ corresponding to particle $n$.}
This way, we can have a $N$-particle operator corresponding to all the elements of the $su(d)$ basis.
For these operators, 
\be\label{eq:secondG}
\sum_{k=1}^{d^2-1} \aver{ G_{k}^2 }_\varrho \leq N (N+d) \Lambda_{\max} 
\ee
holds for all quantum states $\varrho$.

A change of the basis made collectively on all particles corresponds to an orthogonal transformation 
\be\label{eq:trans}
G^\prime_k = \sum_l O_{k l} G_l,
\ee
 with $O=(O^T)^{-1},$ which ensures $\trace(g^\prime_k g^\prime_l)= 2\delta_{kl}.$ 
It is instructive to ask the question how the transformation given in \cref{eq:trans} acts on density matrices.
That is, let us look for $\varrho'$ such that 
\be
\ex{G^\prime_k}_{\varrho'}=\ex{G_k}_{\varrho}\label{eq:condG}
\ee
holds for all $G_k'$ and $G_k.$ It is clear that for $d=2,$ $\varrho'$ can be obtained from $\varrho$ by a unitary rotation.
On the other hand, for $d>2,$ $\varrho'$ cannot always be obtained from $\varrho$ by a physical map, since it can happen that $\varrho$ is physical while
$\varrho'$ fulfilling the conditions given in \cref{eq:condG} for all $k$ is non-physical.

We also define the (symmetric) $N$-particle correlation matrices 
\begin{subequations}\label{eq:collcormat}
\begin{align}
(C_\varrho)_{k l}&:=\frac 1 2 \aver{G_k G_l+G_l G_k}_\varrho  , \label{corrmatrixo} \\
(\gamma_\varrho)_{k l} &:= (C_\varrho)_{k l} -\aver{G_k}_\varrho \aver{G_l}_\varrho  , \label{gammao} \\
(Q_\varrho)_{k l}&:= \frac 1 N \sum_n \left(\frac 1 2 \aver{g_k^{(n)} g_l^{(n)} +g_l^{(n)} g_k^{(n)}}_\varrho-Q_0\delta_{kl}\right), \label{dmatrix} 
\end{align}
\end{subequations}
where we define the constant
\be
Q_0=(d+1) \Lambda_{\max}/(d^2-1)=\frac2 d,
\ee
such that $\Tr(Q_\varrho)=0$ for all states $\varrho$ \footnote{Note that $\aver{g_k^{(n)} g_l^{(n)} +g_l^{(n)} g_k^{(n)}}_\varrho$ is the expectation value of the anticommutator of two $g_k^{(n)}$'s and can be computed from the symmetric tensor $D_{jkl}:=\trace(\{ g_k , g_l \} g_j)$.}. 

Based on the matrices given in Eq.~\eqref{eq:collcormat}, we construct a matrix that plays a central role in our entanglement conditions
\begin{align}
\mathfrak{X}_\varrho&:=(N-1)\gamma_\varrho+C_\varrho-N^2Q_\varrho, \label{xmatrixold}
\end{align}
which has been used in Refs.~\cite{tothPRA09,vitagliano14}. In our paper, we use a somewhat modified definition 
\begin{align}
\mathfrak{U}_\varrho&:=\gamma_\varrho+\frac{1}{N-1}C_\varrho-\frac{N^2}{N-1}(Q_\varrho+Q_0\openone)=\frac1{N-1}\mathfrak{X}_\varrho-\frac{N^2}{N-1}Q_0\openone, \label{xmatrix}
\end{align}
which will turn out to be directly related to the average two-body covariances. All our entanglement conditions will be formulated with $\mathfrak{U}_\varrho$ given in Eq.~\eqref{xmatrix} and $\gamma_\varrho$ defined in Eq.~\eqref{gammao}.

It is interesting to examine, what happens to the matrices given in Eqs.~\eqref{eq:collcormat} and \eqref{xmatrix}, if we change the basis $g_k.$
With a linear transformation of the form \eqref{eq:trans} the correlation matrices transform as 
\begin{equation}
X \mapsto OXO^T
\end{equation}
for all $X\in \{ \gamma, C, Q, \mathfrak{U} \}$.
Furthermore, since these matrices are symmetric it is possible to find an $O$ and a corresponding $G^\prime_k$ such that each of them becomes diagonal. In our paper, we will be interested
mostly in the eigenvalues of $\mathfrak{U}$, and thus its diagonal form, which is then achieved with such an orthogonal transformation. 

We can express the above $N$-particle quantities in terms of correlations on the average $2$-particle density matrix
\begin{equation}
\label{avertwopar}
\rhoavtwo := \frac 1 {N(N-1)} \sum_{i\neq j} \varrho_{ij} ,
\end{equation}
where $\varrho_{ij}$ is the reduced density matrix of the $i$th and $j$th particles. 
Consequently, $\rhoavtwo$ is the reduced two-particle density matrix of the permutationally invariant part of the state \cite{Toth2010Permutationally}
\be
\rhoavtwo=\trace_{N-2} (\varrho_{\rm PI}),
\ee
where the permutationally invariant part of the state is given as 
\be
 \varrho_{\rm PI}=\frac 1 {N!} \sum_{\pi \in \mathcal S_N} V_\pi \varrho V^\dagger_\pi,\label{eq:rhoPI}
\ee
where $\{V_\pi\}$ are the operators corresponding to the $N!$ possible permutations $\pi \in \mathcal S_N$ of the $N$ particles. 
States for which $\varrho_{\rm PI}=\varrho,$ we call {\it permutationally invariant}. They are hence such that
\be
V_\pi \varrho V^\dagger_\pi=\varrho
\ee
holds for any permutation $\pi.$ We recall here that the unitary representation of such a permutation $\pi$ acts on a canonical basis of the $N$-particle state space as follows
\be
V_\pi \ket{j_1,\dots,j_N} = \ket{j_{\pi(1)},\dots , j_{\pi(N)}}.
\ee
Note that for two particles there are only two such permutations, that are represented by the identity $\id$ and the flip operator $F$ respectively.
The flip operator $F$ is the well-known operator exchanging the two parties of a bipartite state. It is defined by the equation
\be
F\ket{\psi_a}\otimes\ket{\psi_b}=\ket{\psi_b}\otimes\ket{\psi_a},
\ee
which holds for any $\ket{\psi_a}$ and $\ket{\psi_b}.$

There is a smaller class of states, called {\it permutationally symmetric} or states with a {\it bosonic symmetry}, for which
 \be
V_\pi \varrho = \varrho V^\dagger_\pi=\varrho
\ee
also holds  for any permutation $\pi.$ For $N$ qubits, pure permutationally symmetric states live in the subspace \btext{spanned} by the symmetric Dicke states given as
\begin{equation}\label{eq:DickeNm}
\ket{{\rm D}_N^{(m)}}=\binom{N}{m}^{-\frac{1}{2}}\sum_k \mathcal{P}_k (\ket{1}^{\otimes m}\otimes\ket{0}^{\otimes (N-m)}),
\end{equation}
where the summation is over all the different permutations of the product state having $m$ particles in the $\ket{1}$ state and $(N-m)$ particles in the $\ket{0}$ state.
For qudits with a higher dimension, we can construct the basis states analogously. Let us see some concrete examples for mixed states of two qubits. The projector onto the singlet state $(\ket{01}-\ket{01})/\sqrt 2$ is permutationally invariant, but it is not permutationally symmetric, while the projector onto the state $(\ket{01}+\ket{01})/\sqrt 2$ is both permutationally invariant and permutationally symmetric.
 
Let us now define the relevant two-body quantities with the average two-particle density matrix. These are
\begin{subequations}\label{eq:Corrsav2}
\begin{eqnarray}
\barCkl&:=&\aver{g_{k} \otimes g_{l}}_{\rhoavtwo}, \label{eq:Cav2} \\
\barXkl&:=&\barCkl - \aver{g_{k}\otimes\openone}\aver{g_{l}\otimes\openone}_{\rhoavtwo}.\label{eq:Xav2}
\end{eqnarray}
\end{subequations}
Note also that the trace of the two-body correlation matrix is related to the flip operator mentioned above
\be\label{eq:flipdef}
F=  \frac 1 d \id \otimes \id + \frac 1 2 \sum_{k=1}^{d^2-1} g_k\otimes g_k
\ee
via the relation
\be
\Tr(\barC)= 2\bigg( \aver{F}_{\rm av2} - \frac 1 d  \bigg) .\label{barC_F}  
\ee

Based on \cref{avertwopar}, simple algebra yields the relation between the $N$-particle expectation value of $G_k$ and the expectation value of $g_k$ on the average two-body matrix as
\be
\aver{G_k}_{\varrho}^2=N^2\aver{g_{k}\otimes \openone}_{\rhoavtwo}^{2}.
\ee
Similarly, the average two-body correlations can be expressed as two-body correlations computed for the two-body density matrix:
\be\label{eq:Cvarrho2Cbar}
(C_\varrho)_{kl} - N (Q_\varrho)_{kl} -NQ_0\delta_{kl}= \sum_{n\neq m} \aver{g^{(n)}_{k} \otimes g_{l}^{(m)}}_{\varrho}=N(N-1)\barCkl.
\ee
Then, using the definition of $\barX$ given in Eq.~\eqref{eq:Xav2} and the definition of $\mathfrak{U}_\varrho$ given in \cref{xmatrix} we obtain
\be\label{eq:Xvarrho2Xbar}
\mathfrak{U}_\varrho = N^2 \barX.
\ee

\subsection{Spin-squeezing polytope of separable states and two-body entanglement}\label{sec:polyspinsqueez}

Here, we summarize the set of spin-squeezing inequalities defined in Refs.~\cite{tothPRA09,vitagliano14}. Consider a system of $N$ spin-$j$ particles.  We define the collective spin components $J_k=\sum_{n=1}^N j_k^{(n)}$, where $k\in\{x,y,z\}$ and $j_k^{(n)}$ are the single-particle spin components of particle $n$.
Note once more that there are only three spin components, which we denote by $J_x, J_y$ and $J_z,$ while there are $(d^2-1)$ $su(d)$ operators, which we denote as $G_1, G_2, ..., G_{d^2-1}.$
In a rotationally-invariant form, the set of generalized spin-squeezing inequalities can be written as \footnote{As we explained earlier, here, with respect to Ref.~\cite{vitagliano14}, we use for convenience a slightly different definition of the $N$-particle correlation matrices. In particular, we consider the matrix $\mathfrak U_\varrho$ instead of $\mathfrak{X}_\varrho$.}
\begin{subequations}\label{eq:spinjsqueezcrit}
\begin{align}
\trace(C_\varrho^{(J)}) &\leq Nj(Nj+1) , \\
\trace(\gamma_\varrho^{(J)}) &\geq Nj , \\
(N-1)\lambda_{\rm min}(\mathfrak U_\varrho^{(J)}) &\geq \trace(C_\varrho^{(J)}) - Nj(Nj+1) , \\
\lambda_{\rm max}(\mathfrak U_\varrho^{(J)}) &\leq \trace(\gamma_\varrho^{(J)}) - Nj   ,
\end{align}
\end{subequations}
where the $N$-particle correlation matrices are defined as 
\begin{subequations}
\begin{align}
    (C_\varrho^{(J)})_{kl} &:=\tfrac 1 2 \aver{J_k J_l + J_l J_k}_\varrho , \\
    (\gamma_\varrho^{(J)})_{kl}&:=(C_\varrho^{(J)})_{kl}-\aver{J_k}_\varrho \aver{J_l}_\varrho , \\
    (Q_\varrho^{(J)})_{kl}&:=\tfrac 1 N \sum_n \left(\tfrac 1 2 \aver{j_k^{(n)}j_l^{(n)}+j_l^{(n)}j_k^{(n)}}_\varrho -Q_0^{(J)} \delta_{kl}\right), \\
    \mathfrak U_\varrho^{(J)}&:=\gamma_\varrho^{(J)}+\tfrac{1}{N-1}C_\varrho^{(J)}-\tfrac{N^2}{N-1}(Q_\varrho^{(J)}+Q_0^{(J)}\openone),
\end{align}
\end{subequations}
i.~e., analogously as in \cref{eq:collcormat}.  In the above criteria, the notation $\lambda_{\rm min/max}(\cdot)$ refers to the minimum/maximum eigenvalue of a matrix. Here again, $Q_\varrho^{(J)}$ is traceless and $Q_0^{(J)}=j(j+1)/3.$

It has been shown in Refs.~\cite{tothPRA09,vitagliano14} that the set of inequalities given in Eq.~\eqref{eq:spinjsqueezcrit} has the properties that 
\begin{itemize}
\item[(i)]  It defines a polytope in the space of $(\aver{\tilde J_x^2}_\varrho,\aver{\tilde J_y^2}_\varrho,\aver{\tilde J_z^2}_\varrho)$ for each fixed value of the mean spin vector $\aver{\vec J}_\varrho=(\aver{J_x}_\varrho,\aver{J_y}_\varrho,\aver{J_z}_\varrho)$.

Here, the ``modified'' second moments are defined as 
\be
\aver{\tilde J_k^2}_\varrho:=\aver{J_k^2}_\varrho-\sum_{n=1}^N \aver{(j_k^{(n)})^2}_\varrho.
\ee
 Thus, they are defined as the global second moments, minus the contribution of the single-particle second moments, this way retaining only the interparticle correlations.
The relation of these quantities with the diagonal elements of $\mathfrak U^{(J)}_\varrho$ is the following:
\be
\frac N {N-1}\aver{\tilde J_k^2}_\varrho - \aver{J_k}_\varrho^2 = (\mathfrak U_\varrho^{(J)})_{kk}.
\ee
\item[(ii)] In the limit $N\gg j$, there exist separable states corresponding to all the extremal points of the polytope. For each fixed value of the vector $\aver{\vec J}_\varrho$, these states are
\be
\begin{aligned}\label{eq:SpinSqueezingextst}
\varrho_{A_k} &= p (\KetBra{\psi_{+,k}})^{\otimes N}+(1-p)(\KetBra{\psi_{-,k}})^{\otimes N} , \\
\varrho_{B_k} &=  \KetBra{\psi_{+,k}}^M \otimes \KetBra{\psi_{-,k}}^{N-M}  ,
\end{aligned}
\ee
where $k=x,y,z.$ Here, $\Ket{\psi_{\pm,x}}$ are the single particle states with  
\be
(\aver{j_x},\aver{j_y},\aver{j_z})=j(\pm c_x,\aver{J_y}_\varrho/Nj, \aver{J_z}_\varrho/Nj)
\ee
and $c_x:=\sqrt{1-(\aver{J_y}_\varrho^2+\aver{J_z}_\varrho^2)/N^2j^2}.$ The other states $\Ket{\psi_{\pm,y}}$ and $\Ket{\psi_{\pm,z}}$ are defined similarly.
Note that the $\varrho_{B_k}$ are pure and non-permutationally-invariant states, while the $\varrho_{A_k}$ are permutationally symmetric (i.~e., bosonic), but mixed.
\item[(iii)] After applying orthogonal transformations $O$ such that $J^\prime_k=O_{kl} J_l,$ we will obtain valid spin-squeezing equations in the new coordinate system. For each quantum state, there is an optimal coordinate system. In the large $N$ limit, the conditions in \cref{eq:spinjsqueezcrit} detect all states that can be detected knowing $C_\varrho^{(J)}, \gamma_\varrho^{(J)},Q_\varrho^{(J)},$ and $\mathfrak U_\varrho^{(J)}.$
\end{itemize}

It can be observed that the spin-squeezing inequalities can be expressed solely in terms of average two-body spin-spin correlations, namely in terms of the quantities $\barCvr^{(J)}$
and $\barX^{(J)}$ analogous to \cref{eq:Corrsav2} but for the spin operators. This suggests that entanglement detected by the set in \cref{eq:spinjsqueezcrit}
could be related to the entanglement of the average two-body reduced density matrix $\rhoavtwo$.
This is indeed the case for permutationally symmetric states, for which \cref{eq:spinjsqueezcrit} detects only states for which the two-body marginal  violates the so-called Computable Cross-Norm and Realignment (CCNR) criterion \cite{Rudolph2003,chen03,Rudolph_2005}, which states that a bipartite separable density matrix $\varrho_{ab}$ must satisfy
\be\label{eq:CCNRcritDef}
\trace|\mathcal{R}(\varrho_{ab})|\leq 1 ,
\ee
where $\mathcal{R}$ is called {\it realigment} map, which can be expressed as $\mathcal{R}(\varrho_{ab})=(\varrho_{ab} F)^{T_b} F$ where again $F$ is the flip operator defined in Eq.~\eqref{eq:flipdef} and $(\cdot)^{T_b}$ is the partial 
transposition acting on party $b$. 

Based on the results of Ref.~\cite{tothguhne09}, it has been shown in Ref.~\cite{tothPRA09,vitagliano14} that, for permutatioanlly symmetric states, violating the CCNR condition is equivalent to violating the Positive Partial Transpose (PPT) condition \cite{peresPPT,Horodecki19961}. The PPT criterion states that any bipartite separable state given as $\varrho_{ab}=\sum_i p_i \varrho_a^{(i)} \otimes \varrho_b^{(i)}$ must satisfy
\be\label{eq:PPTcritDef}
\mathcal T \otimes \idmap (\varrho_{ab}) := \varrho_{ab}^{T_a} \geq 0 , 
\ee
where $\mathcal T \otimes \idmap$ denotes the transposition map acting on the first particle. For bipartite permutationally symmetric states the PPT and CCNR criteria coincide due to the fact that $\varrho_{ab} F=\varrho_{ab}$~\cite{tothguhne09}. 

However, the two entanglement criteria do not coincide with each other for the more general permutationally invariant states, and also for even more general non-permutationally invariant states.
Note also that an $N$-particle state is permutationally symmetric if and only if all of
its two-body marginals are permutationally symmetric. For states that are not permutationally symmetric,  the conditions can detect also states that have separable two-particle reduced states. For further technical details and properties of this set of inequalities we point the reader to Refs.~\cite{tothPRA09,vitagliano14}.

\subsection{Non-physical states within single qudit Bloch ball}\label{sec:nonphysical}

For particles with a dimension $d=2,$ the constraints for positivity of a density matrix are simply given by Eq.~(\ref{tracerho2}), while they are very complicated for $d>2.$ For this reason, we consider a set of states larger than physical states: We consider matrices satisfying the following conditions
\begin{subequations}\label{pseudostate_cond}
\begin{eqnarray}
\varrho&=&\varrho^\dagger,\\
\trace(\varrho)&=&1,\\ 
\trace(\varrho^2)&\le&1,\label{puritycionst}
\end{eqnarray}
\end{subequations}
but we do not require $\varrho\ge0.$ 
From this definition and \cref{eq:absvecg,tracerho2} we see that those state lie within a $d$-dimensional Bloch ball.\footnote{We mention that Ref.~\cite{Krumm2019Quantum} examines quantum computation for the case of Bloch balls with a dimension larger than three.}
In particular, matrices saturating Eq.~\eqref{puritycionst}, which we call pure, are on the boundary of such a higher-dimensional Bloch ball.
As a consequence, every such non-physical quantum state can be given as a mixture of a pure non-physical quantum state and the white noise 
\be
\varrho=(1-p)\varrho_{\rm pure}+p\frac{\id}{d} ,
\ee
where $p$ is a real number that gives the fraction of the white noise component. Here, pure non-physical quantum state can be high-rank states. For instance, let us consider the diagonal density matrix 
\be
\varrho_{\rm pure,\;non-physical}=\frac 2 d\openone - |\Psi\rangle\langle \Psi|,\label{eq:nonphyspure}
\ee
where $\ket{\Psi}$ is a physical pure state.
The state given in \cref{eq:nonphyspure} is non-physical for $d\ge 3,$ it is full rank, and it saturates Eq.~\eqref{puritycionst}, thus it is a pure non-physical state.
Physical states within the higher-dimensional Bloch ball can be rank-2 or having a larger rank, up to full rank,  since now the higher-dimensional Bloch ball contains all physical states, including low rank states. All rank-$1$ physical density matrices are on the surface of the higher-dimensional Bloch ball.

In contrast, note that on the boundary of the set of physical states there are also mixed states. States with a non-full rank density matrix are on the boundary of the set, while states with a full rank density matrix are inside the set.

One can observe yet another interesting property of the non-physical states $\varrho.$
The squared state, $\varrho^2,$ is an unnormalized physical state. Moreover,
if $\varrho$ is a pure non-physical state, then, $\varrho^2$ is a properly normalized physical state (not necessarily pure).
This is related to the well-known purification of density matrices of the type $WW^\dagger,$
where $W$ is not necessarily positive-semi-definite. Now, the purification of the physical state $\varrho^2$ is of the form $\varrho \varrho^\dagger.$

A tensor product of $N$ such non-physical density matrices is
\be
\varrho_{\rm non-phys.\;product} = \varrho^{(1)} \otimes \varrho^{(2)}  \otimes \dots \otimes \varrho^{(N)},\label{eq:aphysprod}
\ee
where $\varrho^{(n)}$ are $d\times d$ matrices that have a unit trace and satisfy \cref{localunc}.
Next, we generalize separable states by defining a wider set, considering 
some mixture of the above product of non-physical states, \eqref{eq:aphysprod}.

\begin{definition}\label{def:1}  We call {\it pseudo-separable} any matrix $\varrho$ that can be decomposed as
\begin{equation}\label{quasisep}
\varrho = \sum_{i} p_{i} \varrho^{(1)}_{i} \otimes \varrho^{(2)}_{i} \otimes \dots \otimes \varrho^{(N)}_{i} ,
\end{equation}
where  $\{ p_i\}$ is a probability distribution, $\varrho_i^{(n)}$ are $d\times d$ matrices that have a unit trace and satisfy Eq.~\eqref{tracerho2}. Then, they also satisfy the uncertainty relation given in Eq.~(\ref{localunc}). Note that we do not require that $\varrho$ or $\varrho_i^{(n)}$ are positive semi-definite. If all $\varrho_i^{(n)}$ are positive semi-definite, then the state given in Eq.~\eqref{quasisep} is a separable state~\cite{Werner1989Quantum}. 
\end{definition} 

\begin{figure}
\includegraphics{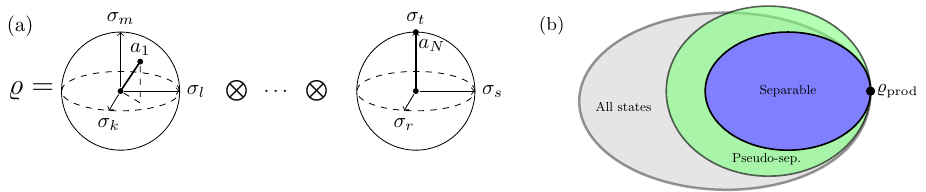}
\caption{(a) A particular pure pseudo-separable state is a product of single-particle matrices with Bloch vectors $\vec a_n$, that are on the boundary of a $(d^2-1)$-dimensional {\it Bloch ball}.
Mixed pseudo-separable states can have shorter Bloch-vector lengths for the single-particle states. Here we show a $3$-dimensional projection. (b) Graphical representation of the space of pseudo-separable matrices with respect to the separable states and the full state space. $\varrho_{\rm prod}=\ketbra{\Psi_{\rm prod}}$ at the boundary of the three sets is a generic pure product state.}\label{fig:pseudosep}
\end{figure}

Clearly, the set of pseudo-separable matrices is convex. Some of these states are not physical, and we will need them to derive our main results.
A subset of pseudo-separable matrices contains physical states. Separable states are part of this set, and in particular all pure product states are at the boundary of the three sets of separable, pseudo-separable and physical states.
A pictorial summary of these statements can be found in \cref{fig:pseudosep}.

While non-physical states are a larger set than physical ones, many relations valid for physical states remain valid also for non-physical states. The length of the generalized Bloch vector is bounded in the same way, as for physical states. Thus, Eq.~\eqref{eq:absvecg} and  Eq.~\eqref{localunc} also hold for non-physical states. Then, for a product of non-physical states, \eqref{eq:aphysprod}, simple algebra yields
\be
(\Delta G_k)_{\varrho_{\rm non-phys.\;product}}^2=\sum_n (\Delta g_k)^2_{\varrho^{(n)}}.\label{eq:sumvar}
\ee
This relation plays a central role in constructing entanglement conditions for physical states.

Let us consider a non-physical state $\varrho$ fulfilling \cref{pseudostate_cond} with the decomposition
\be
\varrho=\sum_i p_i \varrho_i,
\ee
where $p_i$ are probabilities, and $\varrho_i$ are pure non-physical states.
Then, for an observable $A$ we can write
\begin{eqnarray}
(\Delta A)_\varrho^2&=&\sum_i p_i \big\{ \trace(A^2\varrho_i)-\trace(A\varrho_i)^2 + [\trace(A\varrho_i)-\trace(A\varrho)]^2\big\}.
\end{eqnarray}
Hence, it follows that the variance is concave, that is
\be\label{eq:varconcave}
(\Delta A)_\varrho^2\ge \sum_i p_i \left[ \trace(A^2\varrho_i)-\trace(A\varrho_i)^2\right],
\ee
even if $\varrho$ is not a physical state, but just an Hermitean matrix with a unit trace.
Then, many necessary conditions for separability derived with variances of $N$-particle observables also hold for the mixtures of non-physical states described above. It also follows that if we have a concave multivariable function $f(x_1,x_2,...)$ and operators $A_i$ then the expression
\be
f(\ex{A_1}_\varrho,\ex{A_2}_\varrho,...)
\ee
is concave in the non-physical state $\varrho.$

Note that in Ref.~\cite{Toth2006Genuine} there have been non-physical single-qubit states such that
\be
\varrho=\frac{\openone}{2} + \frac 1 2 \sum_{l=1}^3 c_l \sigma_l,
\ee
with $- 1 \le c_l\le 1.$ Here, $\sigma_l$ are the Pauli spin matrices. If all $c_l=1$ then $\va{\sigma_l}=0$ for $l=1,2,3.$ 
It is also worth pointing out that in Refs.~\cite{tothguhne09,Toth2010Separability}, it has been shown that the operator-Schmidt decomposition of bipartite symmetric states is of the form
\be
\sum_k \zeta_k M_k^\prime \otimes M_k^\prime,
\ee
where $\sum_k \zeta_k=1$ and $M_k^\prime$ are pairwise orthonormal observables, i.~e., $\tr(M_k^\prime M_l^\prime)=\delta_{kl}$. For PPT states, we have $\zeta_k \ge 0.$
Thus, all bipartite symmetric PPT states are pseudo-separable, which thus also include physical entangled states.

We can now make the following statement.

\begin{lemma} 
Bipartite pseudo-separable states do not violate the CCNR criterion.  
\end{lemma}

{\it Proof.---}This can be seen as follows. Let us consider the full set of $u(d)$ generators that are not necessarily traceless fulfilling
\be
{\rm Tr}( \tilde g_k \tilde g_l )=2\delta_{kl} \label{eq:ortog2}
\ee
for $k,l=0,1,...,d^2-1.$ For simplifying our derivations later, we number the $\tilde g_k$ from $k=0.$
All states violating the CCNR criterion are detected by an entanglement witness of the type \cite{YuLiu05}
\be
W=\openone\otimes\openone-\frac1 2 \sum_k \tilde g_k^{(1)} \otimes \tilde g_k^{(2)},
\ee
where $\tilde g_k^{(n)}$ is the local $u(d)$ operator $\tilde g_k$ corresponding to particle $n.$
Moreover, for a given $n,$ there are $d^2$ $\tilde g_k^{(n)}$ operators. The expectation value of these operators is related to the purity as
\be
{\rm Tr}[(\varrho^{(n)})^2]=\frac1 2 \sum_k \aver{\tilde g_k^{(n)}}^2_{\varrho^{(n)}},  \label{eq:lambda_purity}
\ee
c.~f. \cref{eq:absvecg}. Let us consider a bipartite product of non-physical states of the form given in \cref{eq:aphysprod}. For this state, based on \cref{eq:lambda_purity},
\begin{align}
\aver{W}&=1-\frac1 2\sum_k \aver{\tilde g_k^{(1)}}_{\varrho^{(1)} } \aver{\tilde g_k^{(2)}}_{\varrho^{(2)} }\ge 1- \frac1 2\sqrt{\sum_k \aver{\tilde g_k^{(1)}}^2_{\varrho^{(1)} } }  \sqrt{\sum_k \aver{\tilde g_k^{(2)}}^2_{\varrho^{(2)} }}\nonumber\\
&= 1-\sqrt{{\rm Tr}[(\varrho^{(1)})^2]{\rm Tr}[(\varrho^{(2)})^2]}\ge0
\end{align}
holds, which is just the same proof that has been given for the product of two physical states  \cite{YuLiu05}. Thus, the product of two non-physical states are not detected by the witness operator. The same holds also for a mixture of products of non-physical states, that is, for pseudo-separable states. \qed

The following bipartite pseudo-separable state with qudits of dimension $d=3$ 
\be\label{eq:entPSsep}
\varrho_{\rm PS3}=\sum_{k=1}^8 \left[(\openone/3+g_k/\sqrt{3})\otimes(\openone/3-g_k/\sqrt{3})+(\openone/3-g_k/\sqrt{3})\otimes(\openone/3+g_k/\sqrt{3})\right],
\ee
where $g_k$ are the Gell-Mann matrices, is  a physical state and it violates the PPT criterion. It is not violating the CCNR criterion, but it is on the boundary of violating it. 

\section{$su(d)$-squeezing entanglement criteria}\label{eq:sudcriteriamain}

We derive here a closed set of inequalities, called the $su(d)$-squeezing inequalities, which are necessarily satisfied by all pseudo-separable states.

\begin{observation}\label{observation:1}
For every given $N$-particle $su(d)$ generators $\{G_k\}_{k=1}^{d^2-1}$, all inequalities in the following set must hold for every $N$-qudit pseudo-separable state $\varrho$
\begin{align}\label{Looind}
\Tr(\gamma_\varrho) - \sum_{k\in I} (\mathfrak{U}_\varrho)_{kk} - 2N(d-1) \geq 0 ,
\end{align}
where $\gamma_\varrho$ and $\mathfrak{U}_\varrho$ are defined by \cref{gammao,xmatrix}, and
$I$ is a subset of $I_{d}:=\{1,2,...,d^2-1\},$ including the case when $I$ is an empty set. 
\end{observation}

{\it Proof.---}Let us consider a pure product of non-physical states of the type given in Eq.~\eqref{eq:aphysprod}. For such states, based on Eqs.~\eqref{eq:sumvar} and \eqref{eq:varconcave}, and based on the uncertainty relation given in Eq.~(\ref{localunc}), it follows that
\be
\Tr(\gamma_\varrho)=\sum_k (\Delta G_{k})_\varrho^{2}= \sum_k \sum_{n=1}^N (\Delta g_{k}^{(n)})_\varrho^{2} \geq 2N(d-1)
\ee
 holds. On the other hand, the diagonal elements of $\mathfrak{U}$ are such that 
 \begin{equation}
 \begin{aligned}
 (\mathfrak{U}_\varrho)_{kk}&=\sum_{n=1}^N (\Delta g_{k}^{(n)})_\varrho^{2}+\frac 1 {N-1}\sum_{n=1}^N \aver{(g_{k}^{(n)})^2}_\varrho
+\frac 1 {N-1}\sum_{n\neq m}^N \aver{g_{k}^{(n)}}_\varrho\aver{g_{k}^{(m)}}_\varrho-\frac N {N-1}\sum_{n=1}^N \aver{(g_{k}^{(n)})^2}_\varrho \\
&=-\sum_{n=1}^N \aver{g_{k}^{(n)}}_\varrho^2+\frac 2 {N-1}\sum_{n=1}^N\sum_{m<n} \aver{g_{k}^{(n)}}_\varrho \aver{g_{k}^{(m)}}_\varrho \leq 0, \label{eq:Xkk}
  \end{aligned}
 \end{equation}
where the inequality comes from the Cauchy-Schwarz inequality. Thus, the diagonal elements of $\mathfrak U$ must be non-positive for product states,
 hence, all such states must satisfy the inequalities
 \begin{align}\label{Looind2}
\Tr(\gamma_\varrho) - \sum_{k\in I}(\mathfrak{U}_\varrho)_{kk} \geq 2 N(d-1)
\end{align}
for all the possible choices of $I \subseteq I_d.$
 
To extend the result to mixtures of product states we use the fact that the left-hand side of Eq.~\eqref{Looind2} is a concave function of the state.
In order to see this, based on Eq.~\eqref{eq:Xkk}, we rewrite left-hand side of Eq.~\eqref{Looind2} as 
 \begin{eqnarray}
\Tr(\gamma_\varrho) - \sum_{k\in I}(\mathfrak{U}_\varrho)_{kk} = \sum_{k\in I_d \backslash I} (\Delta G_{k})_\varrho^{2} -\frac{1}{N-1} \sum_{k\in I_d}  \exs{G_k^2}_\varrho
+\frac N {N-1}\sum_{n=1}^N\aver{(g_{k}^{(n)})^2}_\varrho .\label{eq:gammaXkk}
 \end{eqnarray}
The right-hand side of  Eq.~\eqref{eq:gammaXkk} is clearly concave in the state.
Moreover, Equation~\eqref{Looind2} holds for any choices of the $g_k$ (and thus $G_k$) basis matrices. 
\qed

If we find basis matrices $G_k$ such that $\mathfrak{U}_\varrho$ is diagonal and evaluate \cref{Looind2} with those, then we detect all states detectable with \cref{Looind} and we also need fewer measurements.
In \cref{app:optbasis} (cf. \cref{lemma:1}), we show that we could introduce a further basis operator $g_0$ and consider $u(d)$ generators rather than $su(d)$ generators. However, this would not lead to the detection of more entangled quantum states. Thus, in the rest of the papers, the $g_k$ will be $su(d)$ generators.

Next, based on this optimal basis, we will define a $su(d)$-squeezing parameter.

\begin{observation}\label{observation:2}
Given a density matrix $\varrho$ we define the parameter
\begin{equation}\label{LOOsparam}
\xi_{su(d)}(\varrho) := \Tr(\gamma_\varrho) - 
\sum_{k\in I_+} \lambda_k (\mathfrak{U}_\varrho) -2N(d-1) ,
\end{equation}
where the summation is over all the positive eigenvalues of $\mathfrak{U}_\varrho.$ Then, (i) $\xi_{su(d)}(\varrho) <0$ is a signature of entanglement, while (ii) $\xi_{su(d)}(\varrho) \geq 0$ signals that $\varrho$ cannot be detected with any of the inequalities in \cref{Looind}, in any $N$-particle basis $\{G_k\}_{k=1}^{d^2-1}$. 
\end{observation}
 
{\it Proof.---}We use \cref{observation:1} and define $\xi_{su(d)}(\varrho)$ based on the optimal inequality in Eq.~(\ref{Looind}) that can possibly detect $\varrho$ as entangled. Because there is a minus sign in front of the sum, we need the inequality containing the sum of all the positive diagonal elements of $\mathfrak U_\varrho$. Moreover, it is a consequence of the Schur-Horn theorem that if we order the diagonal elements $(\mathfrak U_\varrho)_{kk}$ of the symmetric matrix $\mathfrak U_\varrho$ in a non-increasing way the following inequality 
\be
\sum_{k=1}^K (\mathfrak U_\varrho)_{kk} \leq \sum_{k=1}^K  \lambda_k (\mathfrak{U}_\varrho) ,  \label{eq:tracegamma}
\ee
holds for all $1\leq K \leq (d^2-1)$.
Thus, the sum of the largest positive elements will be obtained in the basis in which $\mathfrak U_\varrho$ is diagonal and  corresponds to the sum of its positive eigenvalues. Furthermore, a change of orthonormal $N$-particle basis must be of the form \eqref{eq:trans}, which does not change the eigenvalues of $\mathfrak U_\varrho$. Because of that, the minimum on the left-hand side of \cref{LOOsparam} is obtained by summing all and only the positive eigenvalues of $\mathfrak U_\varrho$, i.e., by diagonalizing $\mathfrak U_\varrho$ and taking $I_+$ to be the subset of indices corresponding to the positive eigenvalues. Thus, \cref{LOOsparam} corresponds to the left-hand side of the optimal inequality in the set \cref{Looind} for a given state $\varrho$ and for all possible choices of the $N$-particle basis. \qed 

Based on the properties of the $\mathfrak U$ and straightforward algebra using \cref{eq:secondG} bounding the trace of $C_\varrho,$ the condition in \cref{Looind} of \cref{observation:1} can be expressed as
\begin{align}\label{Looind_alt}
\sum_{k\in I'} (\mathfrak{U}_\varrho)_{kk}+\frac{N(N+d)}{N-1}\Lambda_{\max}-\frac{1}{N-1}{\rm Tr}(C_\varrho) \geq 0 ,
\end{align}
where $I'$ is the complement of the set $I$ such that $I\cup I'=\{1,2,...,d^2-1\}.$ Note that in \cref{Looind_alt} the correlation matrix $C_\varrho$ appears, while in \cref{Looind} one can find the covariance $\gamma_\varrho$ instead. Similarly, the $su(d)$-squeezing parameter \cref{LOOsparam} of \cref{observation:2} can be expressed as
\begin{equation}\label{LOOsparam_alt}
\xi_{su(d)}(\varrho) = \sum_{k\in I_-} \lambda_k (\mathfrak{U}_\varrho) +\frac{N(N+d)}{N-1}\Lambda_{\max}-\frac{1}{N-1}{\rm Tr}(C_\varrho) ,
\end{equation}
where $\lambda_k (\mathfrak{U}_\varrho)$ with $k \in I_-$ are now all and only the negative eigenvalues of $\mathfrak{U}_\varrho$.
We will use these formulas in our derivations.

The quantum state of two qutrit particles given in \cref{eq:entPSsep} is pseudo-separable. Thus, it does not violate the $su(3)$-squeezing inequalities. Simple calculation shows that it saturates the inequality in \cref{Looind} with $I=\emptyset$ for $N=2$ and $d=3,$ and it gives $\xi_{su(d)}(\varrho)=0$. 

\section{Geometrical description and completeness of the set of inequalities}\label{sec:geometry}

In this section, we reformulate the entanglement conditions in \cref{observation:1} in a very simple and transparent form through a geometric picture that generalizes the one introduced in Refs~\cite{tothPRA09,vitagliano14}. This will also serve us to study the question: Does our set of inequalities detect all possible entangled states based on the $N$-particle first and second moments only (as it is the case for the spin-squeezing inequalities of Refs.~\cite{tothPRA09,vitagliano14})?

Here, our answer to this question will be that the inequalities in \cref{observation:1} detect, in the large $N$ limit, all states that can be written as pseudo-separable matrices according to \cref{def:1}. In other words, given definite values of the correlation matrices (\ref{eq:collcormat}) for a large number of particles $N\gg 1$, whenever those values do not lead to a violation of any of the conditions in \cref{observation:1}, then there exists a pseudo-separable matrix that has the same values of the correlation matrices.
Note also once more that knowing the matrices (\ref{eq:collcormat}) is equivalent to knowing the first moments $\aver{G_k}_\varrho,$ and the modified second moments $\aver{\tilde G^2_k}_\varrho$ of all possible $N$-particle observables $G_k$.

\subsection{Setting the space of variables}\label{sec:sudvariablesgeom}

We will use the $N$-particle $su(d)$ generators $\{G_k\}_{k=1}^{d^2-1}$ for our entanglement derivations. 
Then, given a density matrix $\varrho$, in the following we define a set of variables based on our entanglement conditions. For clarity we define them both with average two-body quantities and with $N$-particle quantities (here, we will often omit the label for the density matrix for simplicity, we keep it only for the cases in which some confusion could arise; we remind the reader that throughout this section the density matrix is assumed to be given). First, we define
\be\label{xk} 
x_k= (\mathfrak{U}_\varrho)_{kk}/N^2=(\barX)_{kk} \quad \text{for} \quad k=1,2,...,d^2-1 ,
\ee
which are $d^2-1$ coordinates.
Then, we define [c.~f. \cref{eq:N2}]
\be\label{xd2}
x_{d^2}= \frac 1 {N(N-1)}\left[ \left(N +d \right) \Lambda_{\max}  - \frac 1 {N} \tr(C_\varrho) \right]= \frac 2 N   (1-\aver{F}_{\rm av2})  ,
\ee
which is an additional coordinate, related to how symmetric the density matrix is: $x_{d^2}=0$ corresponds to perfect bosonic symmetry.

The quantities defined above depend on two-body correlations. However, while \eqref{xd2} is linear in the density matrix, Eq.~\eqref{xk} is nonlinear. To make the latter also linear (similarly as done in Refs.~\cite{tothPRA09,vitagliano14}) we fix the components of the vector\footnote{Note again that here we consider only the vector of collective $su(d)$ generators.}
\be
\aver{\vec G}_\varrho=(\aver{G_{1}}_\varrho,\aver{G_{2}}_\varrho,\dots,\aver{G_{d^2-1}}_\varrho).\label{eq:aversigma}
\ee
This way, for each fixed vector $\aver{\vec G}_\varrho$ all $x_k$ are real numbers which are extracted linearly from the quantum state, and which we will use as coordinate of our space later. Then, we construct the vector $(x_1,\dots,x_{d^2})$ and consider the space of such vectors for each fixed value of $\aver{\vec G}$.

Due to the relation between the correlation matrices, the $d^2$ coordinates $x_k$ are constrained by the equation
\begin{align}
\LambdaSigma &= Nx_{d^2} + \sum_{k=1}^{d^2-1} x_k \label{normconst} ,
\end{align}
where the constant on the left-hand side is defined as
\be
\LambdaSigma := \Lambda_{\max}- \sum_{k=1}^{d^2-1}\aver{\vec G}_\varrho^2/N^2,
\ee
i.~e., essentially it contains the norm of the vector of the expectation values of the collective $su(d)$ generators given in \cref{eq:aversigma}.
For fixed $\aver{\vec G}$, we call $\Wspace$ the space with coordinates $(x_1,\dots,x_{d^2})$ such that Eq.~(\ref{normconst}) holds. We emphasize that the definition of such a space depends on the concrete value of $\aver{\vec G}$. In particular that the relation in \cref{normconst} depends on $|\aver{\vec G}|^2$. Hence, that each different value of $|\aver{\vec G}|^2$ gives rise to a different value $\LambdaSigma$.
Note also that there are other constraints coming from physical states, e.~g., $0 \leq \LambdaSigma \leq \Lambda_{\max}$ and $x_{d^2} \geq 0$, which however are not essential for the definition of $\Wspace$. 

\subsection{Representation of the inequalities as polytopes}\label{sec:polytoperep}

Now, we can observe that in the space $\Wspace$ our $su(d)$-squeezing inequalities are represented as a polytope. 

\begin{observation}\label{eq:obs4}
For a given $\aver{\vec G},$ in the space $\Wspace$, the $su(d)$-squeezing inequalities given in \cref{observation:1} are represented simply as the set
\begin{equation}\label{Looind1}
N^2 \left( x_{d^2}+\sum_{k \in I^\prime} x_k \right) \geq 0 ,
\end{equation}
for all possible subsets $I^\prime\subseteq\{1,2,...,d^2- 1\}$. 
\end{observation}

{\it Proof.---}Plugging the definitions of the various variables \eqref{xk} and \eqref{xd2} into the left-hand side of \cref{Looind1} and multiplying by $N^2$ we get 
\be\label{eq:obs4proofeq1}
N^2 \left( x_{d^2}+\sum_{k \in I^\prime} x_k \right) = \frac{1}{(N-1)}\left[ N \left(N +d \right) \Lambda_{\max} -\tr(C) \right] + \sum_{k \in I^\prime} \mathfrak{U}_{kk} . 
\ee
Then, the statement follows from \cref{Looind_alt}. \qed 

The inequalities given in Eq.~\eqref{Looind1} define
half-spaces of $\Wspace$, bounded by hyperplanes defined by the corresponding equality. Here, we are taking into account also the extreme cases $I=\emptyset$ and $I=\{1,2,\dots,d^2-1\}$. Let us try to understand in more detail these hyperplanes.
First of all, we can see that all the hyperplanes would contain the origin, i.~e., the point with 
\be
x_k=0 \quad \text{for}\quad k=1,2,...,d^2. 
\ee
Some of the hyperplanes have more common points with each other. 
However, it is possible to find several groups of $d^2$ hyperplanes, that intersect each other only in the origin. For instance, we can consider the hyperplanes with 
\be
I=\emptyset, \ I=\{1\}, \  I=\{1,2\}, \ \dots, \ I=\{1,2,\dots,d^2-1\}.
\ee
They intersect all together only in the origin, meaning that they define a set of $d^2$ independent hyperplanes. However, the origin lies in $\Wspace$ only whenever $\LambdaSigma=0.$ In this case, 
\be
|\aver{\vec G}|^2=\Lambda_{\max} N^2,
\ee
i.e., the length of $\aver{\vec G}$ is maximal.
Thus, all inequalities are saturated at the origin, which is also a point in $\Wspace$.

\begin{figure}
\begin{center}\includegraphics[width=0.45\linewidth]{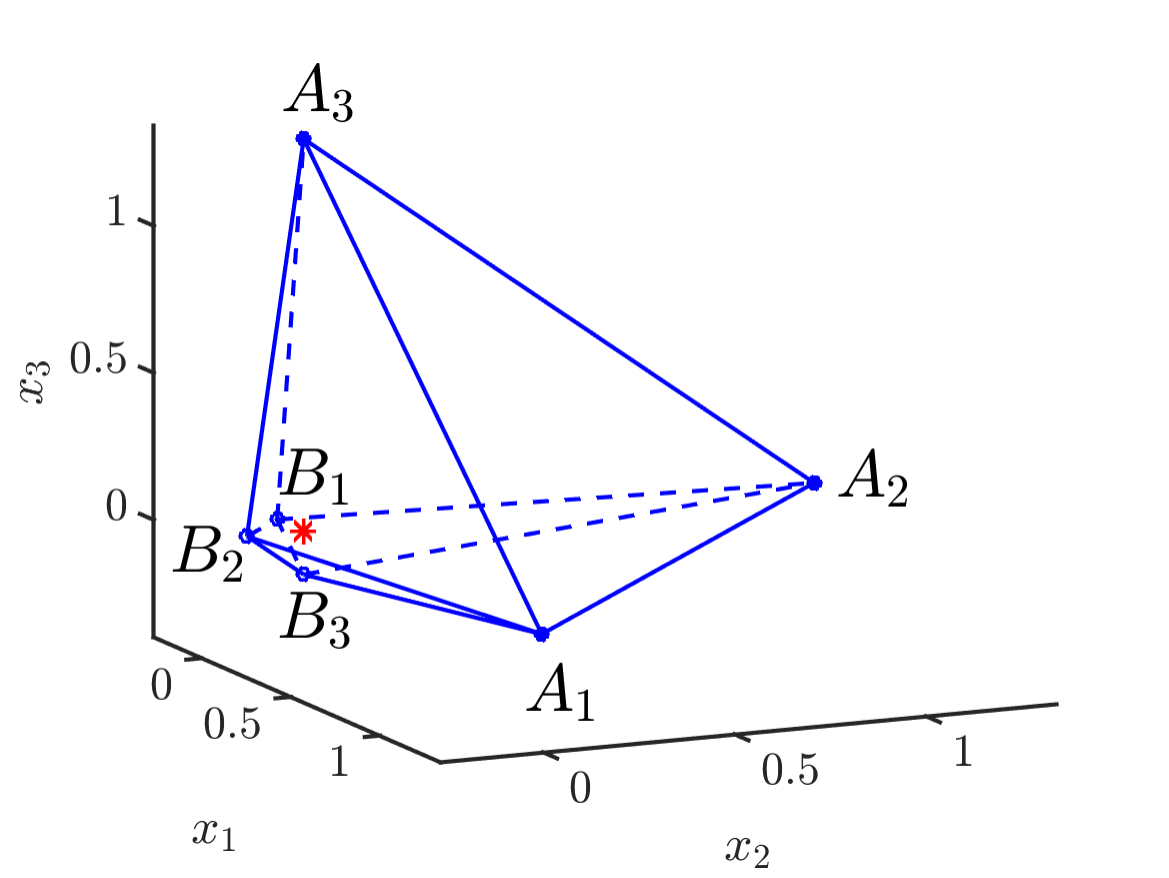}
\includegraphics[width=0.45\linewidth]{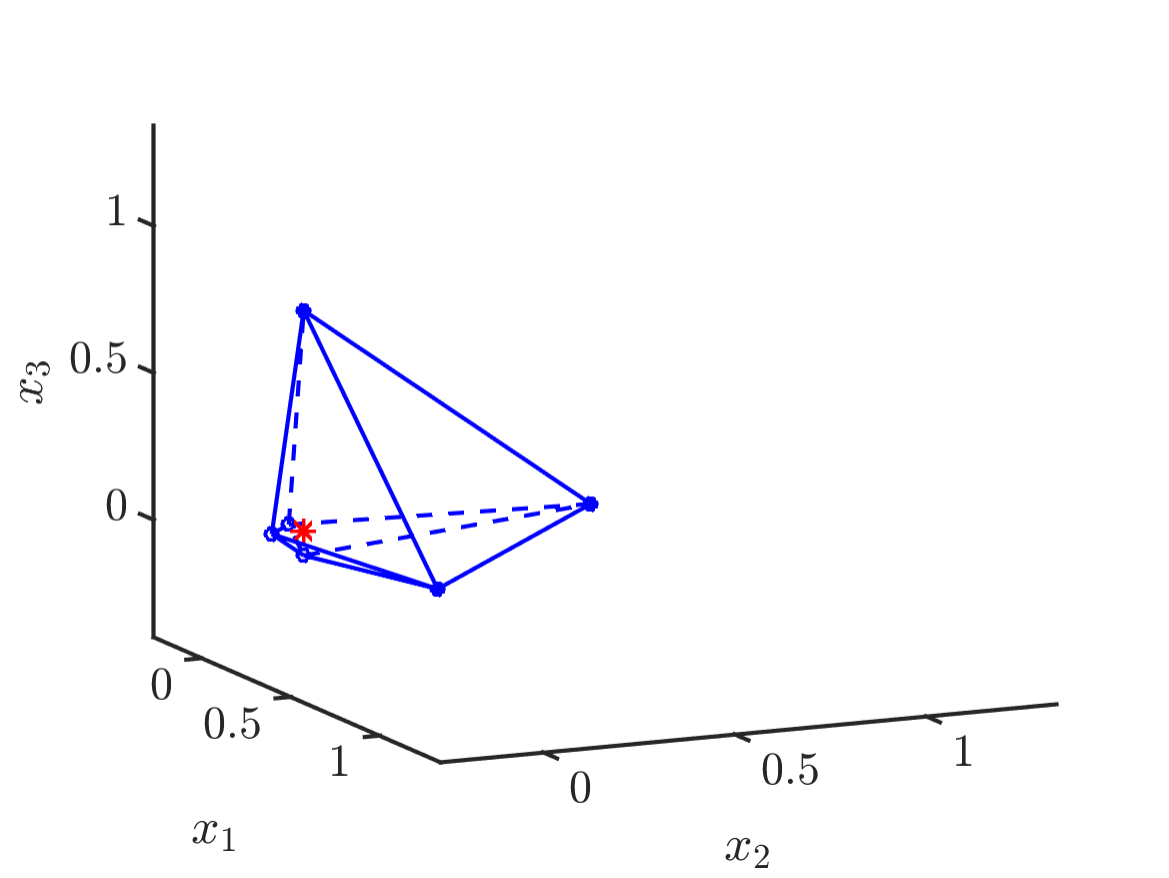}

(a) \hskip6cm (b)

\includegraphics[width=0.45\linewidth]{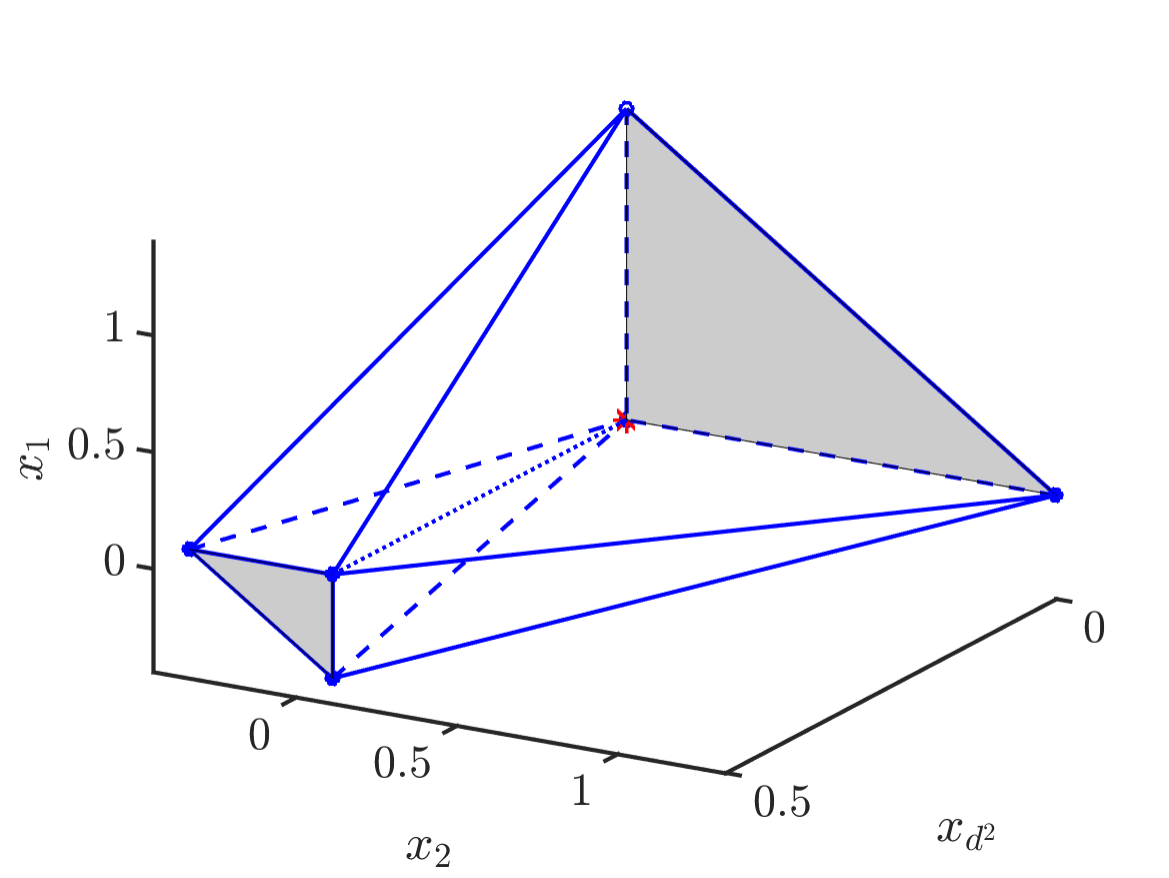}

(c)

\end{center}
\caption{Graphical representation of the $su(d)$-squeezing polytope. The red "*" refers to the origin at $(0,0,0).$ (a) Three-dimensional section of the $d^2$-dimensional polytope obtained for $\LambdaSigma=\Lambda_{\max}$ for $N=10.$ The $x_1,x_2,$ and $x_3$ coordinate axes are shown. In this figure, the $\aver{\vec G}$ vectors are chosen as completely polarized in the last direction, i.~e.,  $\aver{\vec G}^2/N^2=(0,0,0,...,0,\Lambda_{\max} -\LambdaSigma)$.
The two non-intersecting faces are those with vertices $A_1,A_2,A_3$ and $B_1,B_2,B_3$ respectively.
(b) The same for $\LambdaSigma=0.75\Lambda_{\max}.$
(c) Three-dimensional section of  $d^2$-dimensional polytope obtained for $\LambdaSigma=\Lambda_{\max}$ and $N=4.$  The $x_1,x_2,$ and $x_{d^2}$ coordinate axes are shown. 
The dotted line is shown to guide the eye, it is not an edge of the polytope. It is parallel with the $x_{d^2}$ axis and passes through the origin. 
The two non-intersecting faces are filled in gray.}\label{fig:polytope}
\end{figure}

For $\LambdaSigma \neq 0$ there are two hyperplanes from (\ref{Looind1}) that do not intersect each other in $\Wspace$. They are given by 
\be
I_{\rm top}=\emptyset \quad \text{and} \quad I_{\rm bot}=\{1,\dots,d^2-1\}. 
\ee
We can see this as follows. Inserting $I=\emptyset$ in \cref{Looind1} we get that the hyperplane corresponding to $I_{\rm top}$ is given by
\be
x_{d^2}=0 .
\ee
Moreover, from Eq.~(\ref{normconst}), we obtain for the sum of the $x_k$ coordinates from $k=1$ to $k=d^2-1$
\be
\LambdaSigma=\sum_{k=1}^{d^2-1} x_k.
\ee
On the other hand, inserting $I=\{1,\dots,d^2-1\}$ in \cref{Looind1} we find that hyperplane corresponding to $I_{\rm bot}$ is given by
\be
x_{d^2}=-\sum_{k=1}^{d^2-1} x_k , 
\ee
and consequently there is a simple relation between $\Lambda$ and $x_{d^2}$
\be
\LambdaSigma=(N-1) x_{d^2},
\ee
which follows again from \cref{normconst}. See \cref{fig:polytope} for a graphical representation.

In this case, we can observe that, in $\Wspace$, the other hyperplanes from (\ref{Looind1}) intersect both with the hyperplane given by $I_{\rm top}$ and $I_{\rm bot},$ and also intersect with each other, defining the faces and the vertices of the polytope. The vertices are given as
\begin{equation}
\begin{aligned}\label{extrpoints}
(A_k)_{\vec G}&=(\kappa \aver{G_1}^2/N^2,\dots,\kappa c_k^2,\dots,\kappa \aver{G_{d^2-1}}^2/N^2,0) \ ,\\
(B_k)_{\vec G}&=- \tfrac{1}{N-1}(\kappa \aver{G_1}^2/N^2,\dots,\kappa c_k^2,\kappa\aver{G_{d^2-1}}^2/N^2,-\LambdaSigma) \ , 
\end{aligned}
\end{equation}
where we define two constants depending on the expectation values of the collective $su(d)$ generators as
\be
\begin{aligned}
    c_k&=\sqrt{\Lambda_{\max}-\sum_{r\neq k} \aver{G_r}^2/N^2}=\sqrt{\LambdaSigma+\aver{G_k}^2/N^2},\\
    \kappa&={\LambdaSigma}/{\Lambda_{\max}}
\end{aligned}
\ee
for $k=1,2,...,d^2-1.$ Clearly, $0\le\kappa\le1$ holds.
Note once more that in the space we are considering, $\Wspace$, the value of $\aver{G_k}$ is fixed for each $k$, namely they are given constants.

By direct inspection of its coordinates it can be observed that the point $B_k$ is the intersection of all hyperplanes from (\ref{Looind1}) for which $k\in I$ holds, while $A_k$ is the intersection of all hyperplanes for which $k\notin I$ holds.
In particular, the $B_k$ lie all in the hyperplane defined by $I_{\rm bot}$, while the $A_k$ lie all in the hyperplane defined by $I_{\rm top}$.
Thus, the points $A_k$ and $B_k$ for $k=1,2,...,d^2-1$ are the vertices of the polytope defined by Eq.~(\ref{Looind1}) for a fixed $\aver{\vec G}$. We call such a polytope $\OmegaSigma.$ Note, that for $|\aver{\vec G}|^2/N^2=\Lambda_{\max}$ the polytope consists of a single point, namely the origin, and the points $A_k$ and $B_k$ are at $(0,\dots,0)$ for all $k$ in that case. 
On the other hand, for the case $\aver{\vec G}=(0,\dots,0)$ the polytope is the largest.
See Fig.~\ref{fig:polytope} for a graphical representation and \cref{lemma:2} in \cref{app:vertices} for more details.

\subsection{Completeness of the set of inequalities}\label{sec:completenessmain}

Now, we can observe that it is possible to construct $N$-particle matrices corresponding to the vertices $A_k$ as
\be\label{eq:Ykboundarystatesmain}
\varrho_{A_k} =p \varrho^{\otimes N}_{+,k} +(1-p)\varrho^{\otimes N}_{-,k} ,  
\ee
for a certain real number $p,$ which is a function of $\LambdaSigma,$ and a certain single-particle matrices $\varrho_{\pm,k}$. 
See \cref{lemma:3} in \cref{app:vertStat}. 

Moreover, whenever $N_+=Np$ is an integer number, it is also possible to construct $N$-particle matrices corresponding precisely to the $B_k$ as 
\be\label{eq:Xkboundarystatesmain}
\varrho_{B_k} =\varrho^{\otimes N_+}_{+,k} \otimes \varrho^{\otimes (N-N_+)}_{-,k} . 
\ee
When $Np$ is not an integer, one can still take its closest integer $N_+^\prime$ and construct similar matrices $\varrho_{B_k^\prime}$ that correspond to point $B_k^\prime$ such that their distance from the vertices $B_k$ is negligible as compared to the size of the polytope. 
More precisely, we consider the following notion of distance:
\be
\| B_k-B_k^\prime \|=\sum_{i=1}^{d^2} | x_i - y_i | ,
\ee
and find points $B_k^\prime$ that have a distance of order $1/N^2$ from the vertices $B_k$, while the length scales of the polytope, e.~g., the distance between two points $B_k$ and $B_l$ are of order $O(1/N)$. See \cref{lemma:3} in \cref{app:vertStat} for further details. 
The distance that we consider here is directly related to our $su(d)$-squeezing parameter, as we observe later (cf. \cref{eq:sudparsigndist} and \cref{app:distance}).

Note the similarity of the above states with those in Eq.~(\ref{eq:SpinSqueezingextst}) obtained for the spin-squeezing polytope.
There is still one caveat here, however: Here, the single-particle matrices $\varrho_{\pm,k}$ are {\it not necessarily physical density matrices}, i.~e., they can be non-positive operators. Still, they will always satisfy 
\be
\sum_{k=1}^{d^2-1}\trace(\varrho_{\pm,k}g_k)^2=\Lambda_{\max}
\ee
by construction. 
Thus, the matrices in Eqs.~(\ref{eq:Xkboundarystatesmain}) and (\ref{eq:Ykboundarystatesmain}) are pseudo-separable, and they are not always positive.

Given these pseudo-separable matrices corresponding to the vertices, we can then prove that the set $\OmegaSigma$ for all $\aver{\vec G}$ is completely filled by such matrices that are decomposable as in \cref{def:1}. 

\begin{observation}\label{obs:5}
For $N \gg 1$ the set $\OmegaSigma$ for all $\aver{\vec G}$ is filled with points corresponding to pseudo-separable matrices. 
\end{observation}

{\it Proof.---}Let us consider the polytope $\OmegaSigma$ for a certain fixed $\aver{\vec G}$. 
From \cref{lemma:3} in \cref{app:vertStat} we can see that in the limit $N \gg 1$ there exist also pseudo-separable matrices $\varrho_{A_k}$ for every vertices $A_k$ of $\OmegaSigma,$
and they are given in  Eq.~(\ref{eq:Ykboundarystatesmain}).
There exist pseudo-separable matrices $\varrho_{B_k}$ for every vertices $B_k$ of $\OmegaSigma,$
and they are given in  Eq.~(\ref{eq:Xkboundarystatesmain}). 
 Let us then consider a point $P$ internal to the polytope $\OmegaSigma$. Since $\OmegaSigma$ is convex, we can write $P$ as a convex combination of the vertices
\begin{equation}
P = \sum_i p_i V_i,
\end{equation}where $V_i \in \{ A_1,B_1,A_2,B_2,A_3,B_3,\dots \}$
and $\{ p_i\}$ is a probability distribution. Then, let us denote the quantum state corresponding to point $P$  by $\varrho_{P}$
and  the quantum state corresponding to point $V_k$ by $\varrho_{V_k}$.  Then, we have
\begin{equation}
\varrho_{P}=\sum_i p_i\varrho_{V_i},\quad\quad
C_{\varrho_{P}} = \sum_i p_i C_{\varrho_{V_i}}.
\end{equation}
Since $\varrho_{V_k}$ and $\varrho_{P}$ all have the same value for $\aver{\vec G},$ we also have
\begin{equation}
\mathfrak{U}_{\varrho_{P}} = \sum_i p_i \mathfrak{U}_{\varrho_{V_i}},\quad\quad
\gamma_{\varrho_{P}} = \sum_i p_i \gamma_{\varrho_{V_i}}.
\end{equation}
Thus, in the limit $N \gg 1$ every internal point $P$ of $\OmegaSigma$ 
can be mapped to a quantum state $\varrho_{P}$ written as a convex combination of pseudo-separable matrices as in \cref{def:1} and this is true for every fixed value of $\aver{\vec G}$. \qed 

So far, we have fixed the axes of our space in reference to a specific orthogonal basis $\{G_k\}$. In particular, the coordinates $B_k$ with $1\leq k \leq d^2-1$ are just the diagonal elements of the matrix $\mathfrak U$ along some fixed directions given by the $\{G_k\}$. However, ultimately the set of $su(d)$-squeezing inequalities in \cref{Looind} holds for all choices of orthogonal bases $\{G_k\}$. In fact, it can be expressed in a form which is explicitly independent of $N$-particle orthogonal changes of basis as in \cref{eq:trans}, such as the one defining our $su(d)$-squeezing parameter \eqref{LOOsparam}.  

Let us now discuss what happens to the sets $\Omega$ when an orthogonal transformation like the one given in  Eq.~\eqref{eq:trans} is performed. Let us first recall that for every $\aver{\vec G}$ there is a corresponding $\Omega$. The extremal ones are the one for which $\aver{\vec G}=(0,\dots,0)$ and those for which $|\aver{\vec G}|^2/N^2=\Lambda_{\max}$ holds. They represent the largest and the smallest polytopes, respectively.
After a change of the $su(d)$ basis, the first $d^2-1$ axes are ``rotated'' by the the corresponding $N$-particle $O(d^2-1)$ transformation, while the last coordinate $x_{d^2}$ remains fixed.
In the rotated axes, \cref{eq:obs4} still holds, and \cref{Looind1} is the representation of the $su(d)$-squeezing inequalities \eqref{Looind}. 

Given a quantum state $\varrho$, the best way to potentially detect it is to perform an orthogonal $N$-particle change of basis that brings $\mathfrak U_\varrho$ to its diagonal form and thus consider the polytope $\OmegaSigma$ with respect to the corresponding axes and the value of $\aver{\vec G}_\varrho$ taken from the quantum state itself.
In that case, we can represent our $su(d)$-squeezing parameter as 
\be\label{eq:sudparsigndist}
\xi_{su(d)}(\varrho) = N^2 \bigg( x_{d^2}+ \sum_{k \in I_-} x_k \bigg),
\ee
where $I_-$ is the subset of coordinates $x_k$ with $1\leq k \leq d^2-1$ that take a negative value.    
With this notion, each of the $\OmegaSigma$ can be also defined as $\OmegaSigma:\{ P \in \Wspace | \xi_{su(d)}(P)\geq 0 \}$, and in turn $\xi_{su(d)}$ represents a signed distance with respect to $\OmegaSigma$ , see \cref{app:distance} for more details on this point. 
In particular, it is positive when the point $X$ corresponding to the state $\varrho$ lies inside $\OmegaSigma.$ It is zero if $X$ lies on the boundary of $\OmegaSigma$. If the $su(d)$-squeezing parameter $\xi_{su(d)}(\varrho)$ is negative,
then $X$ is outside of $\OmegaSigma$, and $\xi_{su(d)}(\varrho)$ provides the distance from $X$ to the furthest hyperplane to which it is external. 

From the discussion above and the fact that the states in \cref{eq:Xkboundarystatesmain,eq:Ykboundarystatesmain} corresponding to vertices have a diagonal $\mathfrak U,$ which we show in the proof of Lemma~\ref{lemma:3} in Appendix~\ref{app:vertStat}, we can conclude with the following observation. 

\begin{observation}\label{obs:detectall} In the large $N$ limit, the parameter in \cref{LOOsparam} detects all non-pseudo-separable states that can be detected knowing $C, \gamma,Q,$ and $\mathfrak U.$\end{observation}

To conclude this section it is worth noting that for the case $d=2$ the above geometrical construction reduces to the one provided in Ref.~\cite{tothPRA09}, and generalized to higher spin-$j$ systems in Ref.~\cite{vitagliano14}. Thus, in particular, the extremal states corresponding to the vertices of our polytope (and thus also to the internal points) are true separable states in the case $d=2$ and also for higher spin-$j$ operators. For clarity, we also make a more detailed comparison in the following.

\subsection{Relation with Spin-Squeezing polytope}\label{sec:SSpolytopecomp}

Let us see here how the geometric picture provided above generalizes the spin-squeezing polytope discussed in Refs.~\cite{tothPRA09,vitagliano14} and reviewed in Sec.~\ref{sec:polyspinsqueez}. Let us start then, by considering the case $d=2$, where the $su(d)$ operators reduce to the three spin components. First of all, as we recalled in Sec.~\ref{sec:polyspinsqueez}, the spin-squeezing polytope is defined in the space of $(\aver{\tilde J_x^2},\aver{\tilde J_y^2},\aver{\tilde J_z^2},\aver{J_x},\aver{J_y},\aver{J_z})$. More precisely, the polytope can be better visualized in the space of 
\be
\aver{\vec{\tilde K}}:=(\aver{\tilde J_x^2},\aver{\tilde J_y^2},\aver{\tilde J_z^2})
\ee
for each fixed value of the vector 
\be
\aver{\vec J}=(\aver{J_x},\aver{J_y},\aver{J_z}). 
\ee
Then, the completeness property of the set means, loosely speaking, that the spin-squeezing polytope detects all possible entangled states, given information about $\aver{\vec{\tilde K}}$ and $\aver{\vec J}$. 

When employing our $su(d)$-squeezing approach to the $d=2$ case, we again fix the value of the vector $\aver{\vec J}$, but instead of the second moments we consider as coordinates the following quantities
\be
x_k\propto \mathfrak U_{kk} =\aver{\tilde J_k^2}-\aver{J_k}^2=\aver{\tilde J_k^2}-{\rm const.},
\ee and 
\be
x_{d^2} \propto \left[ \left(N +d \right) \Lambda_{\max} -\tfrac 1 {N}\trace(C) \right]=\tfrac 2 N \left[\tfrac{N(N+2)} 4 - |\aver{\vec K}|^2\right],
\ee
where $\aver{\vec K}:=(\aver{J_x^2},\aver{J_y^2},\aver{J_z^2}),$
and we simplified the expressions by not writing the normalization factors explicitly.
Note also that for $d=2$ the modified second moments can be written as $\aver{\vec{\tilde K}}=\aver{\vec K} - N/2$.
 
Thus, here, we extend the geometrical picture in the $d=2$ case from a $3$-dimensional to a $4$-dimensional space for each fixed value of $\aver{\vec J}$. However, these $4$ coordinates obey the relation 
\be
\LambdaSigma=Nx_{4}+x_1+x_2+x_3, 
\ee
and thus we still restrict our problem to a $3$-dimensional subspace. Furthermore, the vertices of our polytope in the $d=2$ case  coincide exactly with the vertices of the spin-squeezing polytope, as in Eq.~(\ref{eq:SpinSqueezingextst}).

For higher dimensions, analogously, we fix the value of $\aver{\vec G}$ and consider the space of 
\be
\aver{\vec{\tilde K}}=(\aver{\tilde G^2_1}, \aver{\tilde G^2_2},\dots,\aver{\tilde G^2_{d^2-1}})
\ee
and the additional coordinate $x_{d^2}$, together with the constraint given in Eq.~\eqref{normconst}.  
The advantage of this is that all our coordinates appear linearly in our parameter, which in turn can be interpreted as a signed distance, as we discussed. 

Still, the vertices of our polytope are a direct generalization of Eq.~(\ref{eq:SpinSqueezingextst}), where, however, the single particle
matrices are just restricted to lie inside a $(d^2-1)$-dimensional Bloch sphere, and thus not necessarily positive.
In fact, in practice most of the vertices correspond to non-physical matrices. For example, in the case $\vec{G}=(0,\dots,0)$ the single-particle components are given by (see \cref{app:vertStat})
\be
\varrho_{\pm,k}:=\tfrac{\id} d \pm \sqrt{\Lambda_{\max}} \ g_k ,\label{eq:rhopm}
\ee
and have a negative eigenvalue for many $su(d)$ generators $g_k$. 

\section{$su(d)$-squeezing and two-body entanglement}
\label{sec:sudsqandtwobody}

Here we show that average two-body correlations provide all information that is needed to compute $\xi_{su(d)}$. 
Afterwards, we analyze whether states detected with $\xi_{su(d)}$ have also entangled two-body marginals. Analogously as in the case of the spin-squeezing inequalities,
we show that this is the case for permutationally symmetric states, but not in the generic case.

\begin{observation}\label{obs:xisudav2}
The parameter $\xi_{su(d)}(\varrho)$ in Eq.~(\ref{LOOsparam}) can be expressed with average two-body correlations as
\begin{equation}\label{xigenav2}
\xi_{su(d)}(\varrho) = N^2 \left( \sum_{k\in I_-} \lambda_k (\barX) + \tfrac 2 N (1-\aver{F}_{\rm av2}) \right) ,
\end{equation}
where 
the summation is over all the negative eigenvalues of $\barX.$ 
\end{observation}

{\it Proof.---}We start from the equation
\be\label{eq:N2}
\frac 1 {N(N-1)}\left[ \left(N +d \right) \Lambda_{\max}  - \frac 1 {N} \tr(C_\varrho) \right]= \frac 2 N   (1-\aver{F}_{\rm av2}),
\ee
which relates the correlation matrix of collective quantities to the average two-particle density matrix.
Equation \eqref{eq:N2} can be verified based on \cref{eq:Cvarrho2Cbar}, which provides a relation between $C_\varrho$ and $C_{\rm av2}$, and  \cref{barC_F}, giving a relation between $\Tr(C_{\rm av2})$ and  $\aver{F}_{\rm av2}$. Beside \cref{eq:N2}, we also need the expression in \cref{eq:Xvarrho2Xbar}, which gives a connection between $\mathfrak{U}_{\varrho}$ and $\barX,$ to show that the right-hand sides of \cref{LOOsparam_alt} and \cref{xigenav2} are equal to each other. \qed 

Thus, our spin squeezing parameter $\xi_{su(d)}(\varrho)$ depends only on the average two-body density matrix $\rhoavtwo.$ We then ask the question, whether it detects only states that have an entangled $\rhoavtwo.$  We show that this is not the case, and present a family of states that are detected by our criterion, while their average two-body density matrix is separable and can even be very close to the completely mixed state.

Let us now look at the definition of our parameter given in \cref{xigenav2}. Our entanglement criterion states that 
\be
\xi_{su(d)}(\varrho) \geq 0 \label{eq:critxi0}
\ee
holds for all $N$-particle separable states. The criterion given in Eq.~\eqref{eq:critxi0} can be rewritten as
\be\label{eq:ourcritlast}
\aver{F}_{\rm av2} \leq 1+ \frac N 2\sum_{k\in I_-} \lambda_k (\barX),
\ee  
where on the right-hand side there is a summation over all the negative eigenvalues of  $\barX.$ 

\begin{observation}
    For permutationally symmetric states, $\xi_{su(d)}(\varrho)<0$ holds if and only if $\rhoavtwo$ is not PPT. 
\end{observation}

{\it Proof.---}Let us consider the parameter given in \cref{xigenav2}. For symmetric multi-qudit states, all two-body reduced states are also symmetric. Thus, $\rhoavtwo$ is also symmetric and hence
\be
\aver{F}_{\rm av2}=1.
\ee
For symmetric states $\varrho_{\rm sym}$, the entanglement criterion given in Eq.~\eqref{eq:ourcritlast} implies that whenever a negative eigenvalue of $\barX$ exists, then the state must be entangled. In particular, in this case the value of the $su(d)$-squeezing parameter is proportional to the sum of the negative eigenvalues of $\barX$, i.~e.:
\be\label{eq:ourcritlast2}
\xi_{su(d)}(\varrho_{\rm sym}) = N^2 \sum_{k\in I_-} \lambda_k (\barX) .
\ee  
Thus, $\xi_{su(d)}(\varrho_{\rm sym})$ is negative if an only if $\barX$ has some negative eigenvalues.
On the other hand, Ref.~\cite{tothguhne09} shows that the correlation matrix of a bipartite symmetric state has negative eigenvalues if and only if the state has a negative partial transpose, i.~e., it violates the criterion defined in \cref{eq:PPTcritDef} and it also violates the CCNR criterion defined in \cref{eq:CCNRcritDef}. \qed 

If a quantum state is not symmetric (i.~e., not bosonic), then
\be
\aver{F}_{\rm av2}<1
\ee
holds, while $\rhoavtwo$ is still permutationally invariant due to its definition in Eq.~\eqref{avertwopar}. Then, the criterion given in Eq.~\eqref{eq:ourcritlast} does also detect only states for which $\barX$ has some negative eigenvalues, however, it does not detect all such states. Moreover, the existence of negative eigenvalues of $\barX$ does not imply that $\rhoavtwo$ violates the PPT criterion for states that are not symmetric~\cite{tothguhne09}. Thus, as we will see in a concrete example, the criterion can detect entangled states for which $\rhoavtwo$ is PPT or even separable. This is possible since, even if $\rhoavtwo$ is separable, there is not an $N$-qubit separable and permutationally invariant state that is compatible with it. 
 
Let us consider now {\it Werner states}, which are states invariant under transformations of the type $U^{\otimes N}$, where $U$ is a single-particle unitary~\cite{Werner1989Quantum}. For two particles, all Werner states can be expressed as a linear combination of the two elements of the permutation group as\footnote{Note that the analogous statement holds more in general for $N$-particle Werner states.}~\cite{Werner1989Quantum}
\be
\varrho_W = \alpha \id + \beta F.\label{eq:Wernerstate}
\ee
Then, the trace of the density matrix is given as
\be
 \tr(\varrho_W) = d^2 \alpha  + d \beta = 1,
 \ee
 and for the expectation value of $F$ we obtain
 \be
 \tr(F \varrho_W) = d \alpha + d^2 \beta ,
 \ee
 where we used that $F^2= \id,$ $\tr(\id)=d^2,$ and $\tr(F)=d$. Based on these, we find that the density matrix can be given as a function of  $\aver{F}_{\varrho_W}$ as
 \be
\varrho_W  = \frac{1}{d^3 - d} \left[ (d-\aver{F}_{\varrho_W}) \id + (d \aver{F}_{\varrho_W} - 1) F \right].
\ee
The Werner state is separable if and only if  \cite{Werner1989Quantum}
\be
\aver{F}_{\varrho_W}\ge 0\label{eq:Wentrcrit}
\ee
holds.
For any traceless $A,$ for the Werner state we have 
\be
\ex{A\otimes A}_{\varrho_W}=\frac{d^3 - d}{d \aver{F}_{\varrho_W} - 1}{\rm Tr}(A^2),\label{eq:corrAA}
\ee
where we used the well-known relation that makes it possible to transform a tensor product into a product of two matrices
\be
{\rm Tr}(F A\otimes A)={\rm Tr}(A^2).
\ee

If the $N$-particle state is a Werner state, so are also its two-body reduced states. Thus, let us consider an average two-body reduced state $\rhoavtwo$ that is a Werner state defined in Eq.~\eqref{eq:Wernerstate}. Let us see how our criterion in \cref{eq:ourcritlast} detects Werner states. 
For Werner states, due to Eq.~\eqref{eq:corrAA}, we have $\aver{g_k \otimes g_k}=\aver{g_l \otimes g_l}$ for all $k, l \geq 1$, i.~e., for all directions of the $(d^2-1)$-dimensional Bloch sphere. Furthermore, we have $\aver{g_k}=0$ for all $k\geq 1$. Thus, we have two cases.
	
  (i) If all eigenvalues $\lambda_k(\barX)$ are non-negative, as we discussed, the state is simply not detected, and in that case we also have
	\be
	\aver{F}_{\rm av2} = \frac 1 d + \frac{1}{2} \sum_{k=1}^{d^2-1}  \aver{g_{k} \otimes g_{k}}_{\rm av2} \geq 0 ,
	\ee
	i.~e., the two-body reduced density matrix is also separable.

(ii) If all eigenvalues $\lambda_k(\barX)$ are negative, then our criterion reduces to
\be
\aver{F}_{\rm av2} \leq 1 + \frac{N}{2} \sum_{k=1}^{d^2-1} \aver{g_{k} \otimes g_{k}}_{\rm av2} = 1 + N \left( \aver{F}_{\rm av2} - \frac 1 d \right) ,
\ee  
which can be rewritten as an inequality bounding $\aver{F}_{\rm av2}$ as
\be\label{eq:ourcritWerner}
\aver{F}_{\rm av2} \geq \frac{N-d}{d(N-1)} . 
\ee
Since for separable Werner states the inequality in Eq.~\eqref{eq:Wentrcrit} holds, we can observe that for $N>d$ our criterion can detect Werner states as entangled that have separable two-body reduced states.

A particular Werner state is given by {\it the $N$-qudit $su(d)$-singlet state} $\varrho_{{\rm singlet}}$, which satisfies 
\be
\aver{G_k^2}=\aver{G_k}=0 , \label{eq:properties}
\ee
for every $N$-particle $su(d)$ generator $G_k$\footnote{Note that due to properties given in Eq.~\eqref{eq:properties}, such a state exists only for $N\geq d$.} and is not a bosonic state. 
From its definition one can immediately obtain that for the reduced two-qudit state of the singlet the expectation value of the flip operator is
\begin{align}
\aver{F}_{\rm av2}&=\frac{N-d^2}{d(N-1)}.\label{eq:Nd2} 
\end{align}
The covariance matrix for the average two-particle matrix is
\begin{align}
\barX&=\diag \left(-\frac 1 d \frac 1 {N-1}, \dots, -\frac 1 d \frac 1 {N-1}\right) ,
\end{align}
Hence, the $su(d)$-squeezing parameter is obtained as
\be
\xi_{su(d)}(\varrho_{{\rm singlet}})=- N(d-1),
\ee
thus the state is detected as entangled.
We can also write down its two-body reduced state
\be
\rhoavtwo=\trace_{N-2}(\varrho_{{\rm singlet}})=\frac{\id \otimes \id}{d^2}-\sum_k \frac 1 {2 d} \frac 1 {N-1} g_k\otimes g_k ,\label{eq:rhoav2_singlet}
\ee
which is a two-particle Werner state. From \cref{eq:Nd2}, we can observe that for $N\ge d^2$ the two-qubit reduced state is separable, since in this case Eq.~\eqref{eq:Wentrcrit} holds. Based on the two-body density matrix given in \cref{eq:rhoav2_singlet}, we can see that the $N$-partite singlet's two-body reduced state tends to the maximally mixed state in the limit $N\rightarrow \infty$. However, the condition in \cref{eq:ourcritlast} detects the state as entangled for all given $N$.

As a measure of robustness, we can consider the mixture of the $su(d)$-singlet state with white noise
\begin{equation}
\varrho_{{\rm noisy\;singlet}}(p_{\rm noise})= (1-p_{\rm noise}) \varrho_{{\rm singlet}} + p_{\rm noise} \frac{\id^{\otimes N}}{d^N},
\end{equation}
 which has the reduced two-qudit density matrix given by
 \be\label{eq:rhoavtwonoisysing}
\rhoavtwo(p_{\rm noise}) = \frac{\id \otimes \id}{d^2}-\sum_k \frac 1 {2d} \frac{(1-p_{\rm noise})}{N-1} g_k\otimes g_k, 
 \ee
which is also a Werner state. The white noise tolerance of our $su(d)$-squeezing parameter in \cref{LOOsparam} is
\be
p_{\rm noise}=1-\frac 1 {d+1}, \label{eq:wnoisetol}
\ee
and is independent of $N$. 

Remarkably, a bipartite maximally entangled state becomes separable exactly after adding the amount of white noise given in \cref{eq:wnoisetol}. However note once more that the state in \cref{eq:rhoavtwonoisysing} is not maximally entangled and actually becomes separable for all $p_{\rm noise}$ for already moderately small values of $N$. To be more specific, let us
see  what the noise level is above which the two-qudit reduced state $\rhoavtwo(p_{\rm noise})$ in \cref{eq:rhoavtwonoisysing} is separable. Calculating the expectation value of the flip operator on such a state we get
\be
  \aver{F}_{\rm av2}=[N-d^2+p_{\rm noise}(d^2-1)]/[d(N-1)].
\ee 
Hence, if $N<d^2$ the noisy singlet's reduced state becomes separable for
 \begin{equation}
 p_{\rm noise}\ge (d^2-N)/(d^2-1),
 \end{equation}
 while if $N\ge d^2$ the singlet has a separable two-qudit density matrix for all $p_{\rm noise}$, which is consistent with \cref{eq:Nd2}.

\section{$su(d)$-squeezing and entanglement at equilibrium}
\label{eq:sudsq}

In this section, we will test our criteria in the equilibrium states of so-called mean-field models \cite{raggio89}, i.~e., in thermal states of the form
\be
\varrho_T= \exp(-H/T)/\trace[\exp(-H/T)] ,
\ee 
for some permutationally invariant Hamiltonian $H$. 

The $d=2$ case of our criteria corresponds to the complete set of optimal spin-squeezing inequalities and the state detected have already been studied in detail in Ref.~\cite{tothPRA09}, where it has been shown that they detect a wider set of states than other similar many-body entanglement witnesses~\cite{stockton03,WangSanders2003,tothpra05,KorbiczEtAl2006,amico08,Guhne2009Entanglement,Ma2011Quantum}. 

Thus, we will focus on $d=3$, which is not covered by optimal spin-squeezing inequalities. 
We can also make a comparison with the spin-$j$ squeezing inequalities of Refs.~\cite{vitagliano11,vitagliano14}, cf. \cref{eq:spinjsqueezcrit}. In particular, note that a spin-squeezing parameter can be defined also for spin-$j$ squeezing criteria 
\begin{equation}\label{xigenav2spin}
\xi_{J}(\varrho) := N^2\left(\sum_{k \in I_-} \lambda_k (\barX^{(J)}) -\frac 1 N [\trace(\barCvr^{(J)}) -j^2] \right) \geq 0 ,
\end{equation}
where again $I_-$ is the set of negative eigenvalues of $\barX^{(J)}$,
with definitions for the matrices $\barX^{(J)}$ and $\barCvr^{(J)}$ similar to \cref{eq:Corrsav2} based on the permutationally invariant two-body reduced states.

\subsection{Thermal states of $su(3)$ models}

As we mentioned, a first important example state detected by our parameter is an $N$-qudit singlet $\varrho_{{\rm singlet}}$. 
Any such singlet state can be obtained as a pure ground state of the Hamiltonian
\begin{equation}\label{eq:Hsudsinglet}
	H_{su(d)} =\frac 1 {N} \sum_{k=1}^{d^2-1} G_k^{2} .
\end{equation}
The equal mixture of all such pure singlets is also a singlet state, obtained as the $T\rightarrow 0$ limit thermal state of the Hamiltonian above.
Such a singlet state 
is also invariant under $U^{\otimes N}$ transformations (i.~e., it is a Werner state) and also under the permutations of the particles,  and it does not lie in the bosonic subspace. 

In that case, there is a  temperature range in which such a thermal state is detected as entangled and is PPT with respect to all bipartitions. This fact can be seen numerically for small $N$ by considering thermal states for the $su(3)$ case, see Table~\ref{tab:sudsinglet}. We can see that the $su(d)$-squeezing criterion can detect PPT entangled states, and that it detects states not detected by the spin-squeezing criterion.

\setlength{\tabcolsep}{5pt}
\begin{table}[hb]
\caption{Limit temperatures for the $su(d)$-squeezing parameter given in \cref{xigenav2}, the spin-squeezing parameter in \cref{xigenav2spin} and the PPT criterion for various various particle numbers for the thermal states of the Hamiltonian given in Eq.~\eqref{eq:Hsudsinglet} for $d=3.$ The limit temperature is the temperature above which these criteria stop detecting entanglement. The PPT criterion detects entanglement if any of the bipartitions is not PPT.} \label{tab:sudsinglet}
\begin{center}
\begin{tabular}{c|c|c|c|c|c}
\hline
\hline
N & 2 & 3 & 4 & 5 & 6 \\
\hline
$T_{su(d)}$ & 2.89 & 3.47 & 3.62 & 3.70 & 3.75 \\
\hline
$T_{\text{spin}}$ & 0 & 1.74 & 1.56 & 1.51 & 1.48 \\
\hline
 $T_{\rm PPT}$ & 5.77 & 4.37 & 3.79 & 3.39 & 3.17 \\
\hline
\hline
\end{tabular}
\end{center}
\end{table}

Other than close to singlet states, we have also tested our $su(d)$-squeezing parameter on thermal states of Hamiltonians of the form
\be\label{eq:Hc}
H(\vec c) = \frac 1 {N} \sum_{k=1}^{d^2-1} c_k G_k^2 , 
\ee 
where $G_k$ are $N$-particle operators constructed from the usual $d$-dimensional generalized Gell-Mann matrices and $c_k$ are random real coefficients. Due to the the prefactor $1/N,$ the various temperature bounds we are interested in will be close to unity for any $N$.
Note that such an Hamiltonian is invariant under permutations of the particles, and so will be its thermal states.

We have found that on average the $su(d)$-squeezing parameter \eqref{xigenav2} detects the thermal state until a higher temperature than the spin-squeezing parameter \eqref{xigenav2spin}, while the PPT criterion detects
the state until an even higher temperature.
 
\subsection{Thermal states of spin-$1$ models}

Here we consider equilibrium states of common (permutationally invariant) spin models, which are also produced in experiments with cold atoms~\cite{pezzerev18}.
In particular, we consider the following Hamiltonian
\be\label{eq:Hspin}
  H_{\rm spin}:= \frac{1} N \left( J_x^2 + J_y^2 + \gamma J_z^2 \right). 
\ee
which can be seen as a mean-field version of the prototypical anti-ferromagnetic XXZ model
and also describes various phases of matter.\footnote{Note that here we use the normalization that is most common in the context of many-body physical models, e.~g., the \btext{Lipkin-Meshkov-Glick} model~\cite{vidal04,vidalpalacios04}.} 
We consider the thermal states of $H_{\rm spin}$ and the two sets of inequalities Eqs.~(\ref{eq:spinjsqueezcrit}) and (\ref{Looind}) with $j=1$ and $d=3$ respectively.
As a result we find that the spin-squeezing inequalities (\ref{xigenav2spin}) detect the state \btext{until a higher temperature \btext{than} the $su(d)$ inequalities given in Eq.~(\ref{xigenav2}).}
Table~\ref{tab:hammodels_1} summarizes our results. 
\begin{table}
\caption{Maximal temperatures until which thermal states of the model given in Eq.~\eqref{eq:Hspin} for $N=6$ and $d=3$ are detected with $\xi_{\rm su(3)}$ and $\xi_{J}$ respectively.}
\label{tab:hammodels_1}
\begin{center}
\begin{tabular}{c|c|c|c}
\hline
\hline
$\gamma$ & 0 & 0.5 & 1.0  \\
\hline
$T_{\text{spin}}$ & 0.53 & 1.02 & 1.28 \\
\hline
$T_{su(d)}$  & 0.26 & 0.56 & 0.71 \\
\hline
\hline
\end{tabular}
\end{center}
\end{table}%

Note that, even if the $su(d)$-squeezing inequalities presented in Eq.~(\ref{Looind}) are a larger set of inequalities than the spin-squeezing inequalities given in 
Eq.~(\ref{eq:spinjsqueezcrit}), they do not detect a strictly larger set of states. The reason is that the completeness of the $su(d)$ inequalities holds only for
pseudo-separable states, while the set of spin-squeezing inequalities is complete for all physical separable states. Thus, there are some quantum states
that are pseudo-separable, and thus not detected by the $su(d)$ inequalities, but still entangled, and possibly detected by the spin-squeezing inequalities.

Next, let us look at the ferromagnetic version of the Hamiltonian given in \cref{eq:Hspin}:
\be\label{eq:Hspin2}
  H_{\rm spin}^\prime:= -\frac{1} N \left( J_x^2 + J_y^2 + \gamma J_z^2 \right). 
\ee
The limit temperatures for the two criteria are shown in Table~\ref{tab:hammodels_2}.  For $\gamma=0,$ the ground state is the symmetric Dicke states of $N$ spin-1 particles with $\ex{J_z}=0.$ Such a Dicke state has the same expectation values for the second moments of the angular momentum components as 
 the symmetric Dicke states of $2N$ spin-{1/2} particles with $\ex{J_z}=0$ described in Eq.~\eqref{eq:DickeNm}
 (e.g., see Appendix~A in \cite{vitagliano14}).
Such Dicke states have a zero expectation value for all spin components and the precision in parameter estimation in linear interferometers using such states can reach, in principle, the maximal Heisenberg-scaling with the number of particles \cite{Lucke2011Twin,Toth2014Quantum,Pezze2018Quantum}. 
In photonic systems, symmetric Dicke states have been created \cite{Kiesel2007Experimental,Wieczorek2009Experimental,Prevedel2009Experimental,Krischek2011Useful,Chiuri2012Experimental}. Symmetric Dicke states have also been created in BECs \cite{Lucke2011Twin,Hamley2012Spin-nematic,Zou2018Beating,Chapman2023Long-Lived}. Such \btext{states} are also useful for 
quantum metrology 
\cite{Lucke2011Twin,Krischek2011Useful,Hyllus2012Fisher,Toth2012Multipartite,Apellaniz2015Detecting,Apellaniz2018Precision,du2025arxiv}. In order to understand the $\gamma=1$ case, we use again that we can straightforwardly map the state of $N$ spin-1 particle to a state of $2N$ spin-1/2 particles. \btext{If $\gamma=1,$
any symmetric $2N$-qubit state is a ground state. Considering the corresponding state of $N$ spin-1 particles, the $T=0$ thermal state is separable.} For other values of $\gamma$, we find that the $su(d)$-squeezing criterion is stronger than the spin-squeezing criterion, as it is summarized in \cref{tab:hammodels_2}.

\begin{table}[h]
\caption{Maximal temperatures until which thermal states of the model given in Eq.~\eqref{eq:Hspin2} for $N=6$ and  $d=3$ are detected with $\xi_{\rm su(3)}$ and $\xi_{J}$ respectively.}
\label{tab:hammodels_2}
\begin{center}
\begin{tabular}{c|c|c|c|c}
\hline
\hline
$\gamma$ & 0 & 0.25 & 0.5 & 0.75 \\
\hline
$T_{\text{spin}}$ & 0.751 & 0.641 & 0.478 & 0.250  \\
\hline
$T_{su(d)}$  & 0.762 & 0.650 & 0.484 & 0.255 \\
\hline
\hline
\end{tabular}
\end{center}
\end{table}%

In conclusion, let us make a comment about the relation of our results to the well-established quantum de Finetti theorems~\cite{STORMER196948,Hudson1976,Fannes1988,CavesFuchsSchack2002,KonigRenner2005,Christandl2007,KonigMitchison09,Trimborn_2016}.
Loosely speaking, they state that permutationally invariant states of many particles, have all few-body reduced states that tend to be separable.
More precisely, permutationally invariant states of $N$ particles with $N\gg 1$ have $k$-body reduced states with ($k\ll N$) such that their distance, as quantified by the trace distance between two density matrices $\tr(|\varrho - \sigma |)$, to the set of separable states is of order $O(k/N)$. In other words, every $k$-body reduced state of an $N$-partite permutationally invariant state can be approximated by a fully separable state, up to an error of order $O(k/N)$ in trace distance.
As we have explained before, our $su(d)$-squeezing parameter is based on just the two-body reduced state of permutationally invariant states (which corresponds to the quantum state averaged over all particle permutations). Due to the de Finetti theorem then, such a $2$-body reduced state state will be approximately separable, up to an error which scales as $1/N$ in trace distance. Nevertheless, even if the reduced state is very close to separable, or even separable, it can be incompatible with a globally separable (and permutationally-invariant) state. We have shown that it is possible to prove such a property in some cases simply from our $su(d)$-squeezing parameter. In particular, we have seen in the exemplary class of Werner states such as the $su(d)$-singlet, which are also a complex and interesting class of states on their own~\cite{Werner1989Quantum,PhysRevA.64.062307,EggelingWerner01,Christandl2007,HuberKlepMagronVolcic2022}. 
Furthermore, our conditions do not require the state to be permutationally invariant, but it just uses information from its permutationally invariant part.
Therefore, generic states can also be detected even when its average two-body reduced state is separable.

\begin{figure}
\begin{center}\includegraphics[width=0.45\linewidth]{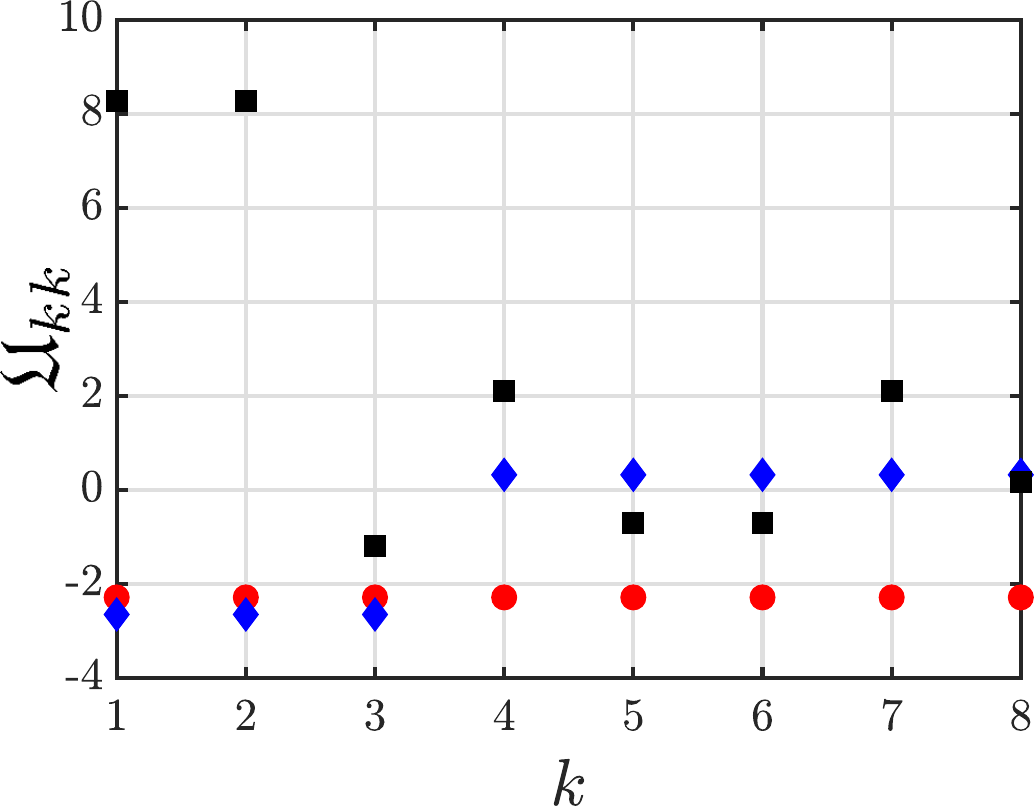}
\end{center}
\caption{The diagonal elements $\mathfrak U_{kk}$ for the thermal states of (circle) the Hamiltonian given in Eq.~\eqref{eq:Hsudsinglet} for $T=1,$ (diamond) the Hamiltonian given in Eq.~\eqref{eq:Hspin} for $\gamma=1$ and $T=0.5,$ and (square) the Hamiltonian given in Eq.~\eqref{eq:Hspin2} for $\gamma=0$ and $T=0.5,$ where $N=4$ and $d=3.$ Note that we use the local $su(d)$ generators given in Eq.~\eqref{eq:gk}, thus all these states have a diagonal $\mathfrak U.$}\label{fig:Ueig}
\end{figure}

The thermal states of the Hamiltonians considered in this section are compared in Fig.~\ref{fig:Ueig}. We choose 
\be
\{g_k\}_{k=1}^8=\{j_x,j_y,j_z,\{j_x,j_y\},\{j_y,j_z\},\{j_x,j_z\},j_x^2-j_y^2,\sqrt{3}j_z^2-2/\sqrt{3}\openone\},\label{eq:gk}
\ee
where $j_l$ the angular momentum components for $d=3$
and $\{A,B\}=AB+BA$ is the anticommutator.  Here, based on Eq.~\eqref{eq:Xvarrho2Xbar}, we know that
\be
\mathfrak U_{kk}\propto \ex{g_k \otimes g_k}_{\rm av2}-\ex{g_k \otimes \openone}^2_{\rm av2}.
\ee
For the thermal state of the Hamiltonian given in Eq.~\eqref{eq:Hsudsinglet}, which is close to an $su(d)$ singlet, we have an anticorrelations for all $k.$
For the thermal state of the Hamiltonian given in Eq.~\eqref{eq:Hspin} with $\gamma=1$, which is close to an $su(2)$ singlet,  we have anticorrelations for $k=1,2,3.$ 
Finally, for the thermal state of the Hamiltonian given in Eq.~\eqref{eq:Hspin2} with $\gamma=0$, which is in the vicinity of a symmetric Dicke state of $N$ spin-$1$ particles with $\ex{J_z}=0,$ we have strong correlations for $k=1,2.$ 
Such a state can be obtained from the symmetric Dicke state of $2N$ qubits [see the $N$-qubit Dicke state in Eq.~\eqref{eq:DickeNm}] mapping each qubit pair to a spin-$1$ particle using $|11\rangle\rightarrow|+1\rangle,(|01\rangle+|10\rangle)/\sqrt{2}\rightarrow|0\rangle,$ and $|00\rangle\rightarrow|-1\rangle$
.

\section{Conclusions and outlook}\label{sec:conclusions}

We have derived \btext{a} set of $su(d)$-squeezing inequalities and defined a corresponding $su(d)$-squeezing parameter that detects entangled multi-qudit states that cannot be written in a certain pseudo-separable form. Our conditions are formulated with the first and second moments of collective operators that are the sums of single-particle $su(d)$ generators, however, they do not depend on the particular choice of the $su(d)$ generators, \btext{which can be chosen optimally for a given known quantum state in a straightforward way, i.e., as the principal directions in which our collective correlation matrix $\mathfrak U$ is diagonal}. We have also provided a geometric interpretation of the parameter as a signed distance from the set of pseudo-separable states, which forms a convex polytope in the space of collective $su(d)$ averages and correlations.
We studied how the parameter detects thermal states of spin-models and $su(d)$-models with a permutation symmetry \btext{and} obtained several interesting results. (i) States close to $su(d)$ singlets are detected as entangled, \btext{some of} which are also PPT with respect to all possible bipartitions. (ii) Permutationally symmetric states are detected if and only if their average two-body reduced state is not PPT. (iii) The comparison with spin-squeezing inequalities seems to be nontrivial. One can find example states that are detected by either set of inequalities and not by the other, at least for small values of $d$ and $N$.  The question of what happens in the large $N$ limit is still open and, also in the light of our results, might be nontrivial.

\btext{Furthermore, in the light of our results it might be interesting to look for a similar approach involving different sets of operators, and understand, for example, 
what the minimal set of inequalities is that characterizes all possible separable states (i.e., not just pseudo-separable as in our work) that can be described in terms of just first and second moments of collective operators. Likely such a minimal set of conditions will contain $su(d)$ generators as well as smaller sets of observables, like $su(2)$ generators in the $d$-dimensional representation.}

\btext{Our} results suggest a possible application of the framework to a deeper study of entanglement in 
many-body $su(d)$ models, especially mean-field models \cite{raggio89} and a deeper exploration of potential connections with the seminal results in that framework, e.~g., de Finetti theorems~\cite{Christandl2007,KonigMitchison09}. Experimentally, being based on just bipartite correlations these criteria could find \btext{applications} to detect entanglement in a wide range of systems, especially in many-body scenarios such as atomic clouds and condensed matter systems~\cite{Guhne2009Entanglement,FriisNatPhys19,FrerotFadelLewenstein23}. \btext{It could be used to to detect entanglement of the spin-1 singlet created in atomic ensembles \cite{behbood14}, or to detect the entanglement of spin-1 Dicke states created in BEC's \cite{Zou2018Beating}, or detecting entanglement in experiments with large spins on an optical lattice \cite{Alaoui2022Measuring}. }
In the future it would be interesting to compare these sets of criteria with other approaches based on two-body correlations and uncertainty relations~\cite{hofman03}, such as the well-known covariance matrix criterion \cite{guhnecova,gittsovich08,PhysRevA.82.032306}. Such criteria have been also recently employed to quantify entanglement via so-called entanglement monotones~\cite{FadelVitagliano_2021,liu2022bounding}.
In the light of our results as compared to \cite{vitagliano14}, and also in the light of recent studies with a similar approach~\cite{M_ller_Rigat_2022} 
it would be particularly interesting to study what is the relation between the specific observables considered for entanglement detection and the symmetries of the states detected with the various criteria.
Furthermore, one could extend the framework to distinguish different levels of multipartite entanglement, such as the so-called depth of entanglement~\cite{Lucke2014Detecting,vitagliano16,PhysRevA.97.020301}, and to include higher order correlations of the observables.  The details of the numerical calculations
are discussed in \cref{app:nuemrics}

\begin{acknowledgments}
We thank Marcus Huber, Jens Siewert and Reinhard Werner for discussions. We thank M. Navascues for drawing our attention to Ref.~\cite{Krumm2019Quantum}. GV acknowledges financial support from the Austrian Science Fund (FWF) through the grants P 35810-N and P 36633-N (Stand-Alone).
We acknowledge the support of the  EU (QuantERA MENTA, QuantERA QuSiED, COST Action CA23115),
the Spanish MCIU (Grant No.~PCI2022-132947), the Basque Government (Grant No. IT1470-22), and the National Research, Development and Innovation Office of Hungary (NKFIH) (2019-2.1.7-ERA-NET-2021-00036). We thank the National Research, Development and Innovation Office of Hungary (NKFIH) within the Quantum Information National Laboratory of Hungary.   We acknowledge the support of the Grant No.~PID2021-126273NB-I00 funded by MCIN/AEI/10.13039/501100011033 and by ``ERDF A way of making Europe''.  We thank the ``Frontline'' Research Excellence Programme of the NKFIH (Grant No. KKP133827). We thank Project no. TKP2021-NVA-04, which has been implemented with the support provided by the Ministry of Innovation and Technology of Hungary from the National Research, Development and Innovation Fund, financed under the TKP2021-NVA funding scheme. G.~T. is thankful for a  Bessel Research Award from the Humboldt Foundation.
O.G. acknowledges the Deutsche Forschungsgemeinschaft (DFG, German Research Foundation, project numbers 447948357 and 440958198), the Sino-German Center for Research Promotion (Project M-0294) and the German Ministry of Education and Research (Project QuKuK, BMBF Grant No. 16KIS1618K).
\end{acknowledgments}

\appendix 

\section{More general $N$-particle basis}\label{app:optbasis}

Here \btext{briefly reformulate our conditions in a bit more general way, by considering a full-basis of the single-particle operators} 
\btext{Thus,} instead of the traceless $su(d)$ generators introduced in Sec.~\ref{sec:collcovs}, we consider a full set of $u(d)$ generators $\tilde g_k$ given at the end of \btext{Sec.~\ref{sec:nonphysical} that are not necessarily traceless, but are still mutually orthogonal.}
In the new basis, we can derive \btext{similar} entanglement criteria as in \cref{observation:1}.
We show \btext{here} that these entanglement criteria do not detect more entangled states.

In particular, instead of the expansion in \cref{eq:rhodef} one could consider a more general basis $\tilde g_k$ such that 
\be
\varrho = \frac1 2\sum_{k=0}^{d^2-1} \aver{\tilde g_k}_\varrho \tilde g_k,
\ee
\btext{where for the generators \cref{eq:ortog2}
holds.}
For simplifying our derivation, we number the $\tilde g_k$ from $k=0.$ 
Let us now consider a larger set of \btext{collective correlation matrices as in \cref{eq:collcormat} as well as a} matrix $\tilde{\mathfrak{U}}_\varrho$ defined analogously as  in \cref{xmatrix},
such that there is also an index $k=0.$

\btext{It is easy to see that following the same proof as in \cref{observation:1} we can prove that all inequalities in the following set
\begin{align}\label{Looindapp}
\Tr(\tilde \gamma_\varrho) - \sum_{k\in I} (\tilde{\mathfrak{U}}_\varrho)_{kk} - 2N(d-1) \geq 0 ,
\end{align}
also hold for every $N$-qudit pseudo-separable state $\varrho$. 
Note that the bound $2N(d-1)$ comes once again directly from the purity bound, that in this case simply reads
\be
\tr(\varrho^2) = \sum_{k=0}^{d^2-1} \aver{\tilde g_k}^2_\varrho \leq 2 ,
\ee
and leads to the same bound for the sum of the variances as in \cref{localunc}:
\be
\sum_k (\Delta \tilde g_k)^2_\varrho \geq 2 (d-1).
\ee
}

Below, we prove that such an extended \btext{formulation} does not bring essentially any advantage for detecting entanglement via our method.
\begin{lemma}\label{lemma:1}
For every state $\varrho$ there exists always a basis $\tilde g_k$ with the following characteristics. First, \btext{$\tilde g_0=\sqrt{\frac{2}{d}}\id.$} Second, $\tilde g_k$ for $k=1,2,...,d^2-1$ are the $su(d)$ basis operators, 
such that $\tilde{\mathfrak{U}}_\varrho$ is diagonal using that basis. 
\end{lemma}

{\it Proof.---}Let us consider a general state $\varrho$ and express $\tilde{\mathfrak{U}}_\varrho$ in $N$-particle basis of the type mentioned above, with 
\be
\tilde G_0=N\sqrt{\frac{2}{d}}\id_N ,\label{eq:G0}
\ee
which is the collective operator corresponding to $\tilde g_0.$ The key point is to observe that for any state $\varrho$ we have $(\tilde{\mathfrak{U}}_\varrho)_{0,k}=0$ for all $k=0,\dots,d^2-1$. In fact, we have
\be
(\tilde{\mathfrak{U}}_\varrho)_{0,k}=N\left(\left\langle \tilde G_0  \cdot \tilde  G_k\right\rangle_\varrho-\sum_n \left\langle \tilde  g_0^{(n)}  \cdot \tilde  g_k^{(n)}\right\rangle_\varrho \right)-(N-1)\left\langle \tilde G_0\right\rangle_\varrho \left\langle \tilde  G_k\right\rangle_\varrho.
\ee
That is, we have
\be
(\tilde{\mathfrak{U}}_\varrho)_{0,k}= N\sum_{n\neq m}\left\langle \tilde  g_0^{(n)} \otimes \tilde g_k^{(m)}\right\rangle_\varrho-(N-1)\left\langle \sum_n \tilde g_0^{(n)} \right\rangle_\varrho\left\langle\sum_n \tilde  g_k^{(n)}\right\rangle_\varrho=0
\ee
independently from $\varrho$. Thus, in such basis $\mathfrak{U}_\varrho$ is of the form
\begin{equation}
\tilde{\mathfrak{U}}_\varrho=\left(
\begin{array}{cc}
 0 & 0 \\
 0 & \mathfrak{U}_\varrho
\end{array}
\right) \ .
\end{equation}
The final diagonal form is obtained diagonalizing $\mathfrak{U}_\varrho$, i.~e., finding the optimal $su(d)$ basis. \qed   

Thus, the eigenvalues of $\tilde{\mathfrak{U}}_\varrho$ will be the same as the eigenvalues of $\mathfrak{U}_\varrho$ apart from an additional zero eigenvalue. 
\btext{Moreover, the additional term in $\tr(\tilde \gamma)$ is also equal to zero in that basis. Thus,}
since the entanglement criterion in \cref{observation:1} is based on the sum of the negative eigenvalues, we would not detect more entangled states with the extended matrix.

\btext{As a final comment, we note that one could also think to consider an overcomplete observable basis, in which the operators are not necessarily orthogonal and still use our approach to derive entanglement conditions, which are essentially based on the bound on the single-particle purity. 
Let us consider such a set of operators $\tilde g_k^\prime$ and first its covariance matrix on a single-particle state, namely the matrix
\be
[\gamma]_{kl} = \tfrac{1}{2} \aver{\tilde g_k^\prime \tilde g_l^\prime + \tilde g_l^\prime \tilde g_k^\prime} - \aver{\tilde g_k^\prime} \aver{\tilde g_l^\prime} ,
\ee 
calculated on single-particle states.
Note that $\gamma$ is still a symmetric matrix, but now has a dimension potentially larger than $d^2$.
Nevertheless, it can be diagonalized by an orthogonal transformation, i.e., $\gamma^\prime = O\gamma O^T$ that effectively maps the observable basis $\{ \tilde g_k^\prime \}$ into an orthonormal one $\{ \tilde g_k \}$ in a 
$d^2$-dimensional subset.
Because of that the diagonalized covariance matrix will have the form $\gamma^\prime = \gamma_{su(d)} \oplus 0$, i.e., it will have a larger kernel, but the same number of nonzero diagonal elements, that are still obtained as variances of
$su(d)$ generators. Thus we still obtain
\be
\Tr(\gamma) = \sum_k (\Delta \tilde g_k^\prime)^2 = \sum_k (\Delta \tilde g_k)^2 \geq d-1 .
\ee
Thus, our method to derive the inequalities as in \cref{observation:1} still applies equally, and one would just get a larger set of inequalities. 
However, following the same argument as above we can see that the eigenvalues of the corresponding matrix $\tilde{\mathfrak{U}}_\varrho$ (which is still a symmetric matrix) are still obtained in an orthonormal basis of the form 
as in \cref{lemma:1}, i.~e., composed by the identity matrix plus an $su(d)$ basis. 
Thus, the form of our criteria as in \cref{observation:1} is in that sense still optimal.
}

\section{Vertices of $\OmegaSigma$}\label{app:vertices}

\begin{lemma}\label{lemma:2}
For every \btext{$\aver{\vec G}$}, the set $\OmegaSigma$ has vertices $A_k$ and $B_k$ as given in Eq.~(\ref{extrpoints}). 
\end{lemma}

{\it Proof.---} Let us consider the points $A_k$ and $B_k$ as in Eq.~(\ref{extrpoints}) for a fixed \btext{$\aver{\vec G}$}. For each choice of $I$, Eq.~(\ref{Looind1}) defines a linear half space, the boundary being a hyperplane. Then, at first we can observe that the points $A_k$ and $B_k$ lie in two ``extremal'' hyperplanes, obtained with $I_{\rm bot}=\{1,\dots,d^2-1\}$ and $I_{\rm top}=\emptyset$ respectively, namely 
\begin{subequations}\label{eq:planesboth}
\begin{align}
x_{d^2} &= -\sum_{r=1}^{d^2-1} x_r \label{tracegammain} , \\
x_{d^2} &= 0 \label{tracecin} .
\end{align}
\end{subequations}
Let us then consider the subspace $\Wspace$ defined by Eq.~(\ref{normconst}).
The intersection of Eqs.~(\ref{tracecin}) and \eqref{tracegammain}, together with the constraint (\ref{normconst}), is the subspace defined by $\sum_{k=1}^{d^2-1} x_k=x_{d^2}=\LambdaSigma=0$, that contains the origin. In particular, the origin $A_k=B_k=0$ defines the whole polytope $\OmegaSigma$ for every $\aver{\vec G}$ such that $|\aver{\vec G}|^2/N^2=2\tfrac{d-1} d$. 

For $\LambdaSigma\neq 0$, and taking also account of the constraint (\ref{normconst}), (\ref{tracecin}) and (\ref{tracegammain}) cannot be saturated together.
Then, substituting the coordinates (\ref{extrpoints}) into Eq.~(\ref{Looind1}) we find that the points $B_k$ correspond to the intersections of all hyperplanes
$x_{d^2} +\sum_{r\in I} x_r = 0$,
such that $k \in I$. Conversely, the points $A_k$ correspond to the intersections of all such hyperplanes with $k \notin I$. Thus, we have $d^2-1$ points corresponding to the intersection of all inequalities but Eq.~(\ref{tracecin}). 
These are the $B_k$. And also, there are $d^2-1$ points corresponding to the intersection of all inequalities but Eq.~(\ref{tracegammain}). These are the $A_k$. 

On the other hand, the boundary of Eq.~(\ref{Looind1}) for every subset of indices $I$ can be defined as the subspace that contains all the vertices $B_k$ such that $k\in I$ and all the vertices $A_k$ such that $k\notin I$. Thus, the boundary of every inequality in the set (\ref{Looind1}) corresponds to a hyperplane defined by a subset of the vertices (\ref{extrpoints}). Hence, the two objects defined by respectively the vertices $B_k$ and $A_k,$ and by the boundary planes of the set (\ref{Looind1}) are equivalent for every $\aver{\vec G}$. \qed 

\section{$\xi_{su(d)}$ as a signed distance in $\Wspace$}\label{app:distance}

We will relate $\xi_{su(d)}$ to the signed distance of the point corresponding to the quantum state from the polytope described by $\OmegaSigma.$
As a signed distance between two points $X$ and $Y$ we consider
\be
S_D(X,Y) = \sum_{i=0}^{d^2} (x_{i}-y_i),
\ee
which is obtained from $\| X-Y \|=\sum_i | x_i - y_i |$ by removing the absolute value $| \cdot |.$ 

Let us now assume that $x_i$ are the coordinates of a general point $X$ corresponding to a given quantum state and for that state we have the expectation values of the $N$-particle operators $G_k$ given in Eq.~\eqref{eq:Sigmak}. Then, let us assume that $A_k$ are the coordinates of a point 
saturating one of the inequalities given in Eq.~\eqref{Looind1} for some $I,$  where $I$ is a certain subset of the coordinates  with $1\leq k\leq d^2-1.$ Hence
\be
y_{d^2}+\sum_{l\in I} y_l= 0\label{eq:facet}
\ee
holds for a given set $I.$ The equation in Eq.~\eqref{eq:facet} corresponds to hyperplane that contains a face of the polytope $\OmegaSigma$ corresponding to a given $\langle \vec G \rangle.$ Then, for the point closest to $X$ on that hyperplane
\be
A_k=B_k
\ee
 holds for all $k\notin I.$ Thus, the signed distance of $X$ from the hyperplane corresponding to the set $I$ is 
\be
S^{I}_D(X)=x_{d^2}+\sum_{i\in I} x_i.
\ee
Then, our parameter $\xi_{su(d)}(\varrho)$ gives the minimal signed distance from the point $X,$ corresponding to $\varrho,$ to the boundary of $\OmegaSigma$, namely $\xi_{su(d)} (\varrho)/N^2= \min_{I} S^{I}_D(X)$,
where the minimization is over all possible  $I$ corresponding to all the different hyperplanes.
 
\section{Quantum states at the vertices of the polytope}\label{app:vertStat}

\begin{lemma}\label{lemma:3}
Let us consider a fixed value of $\aver{\vec G}$ and the corresponding $\OmegaSigma$. In the limit $N \gg 1$, there exist pseudo-separable matrices $\varrho_{A_k}$ and $\varrho_{B_k}$ corresponding to the vertices $A_k$ and $B_k$, respectively, and this holds for every given $\aver{\vec G}$. 
\end{lemma}

{\it Proof.---} Let us consider an $su(d)$ basis $\{ g_r\}$ and the following matrices
\begin{equation}
\begin{aligned}\label{eq:Fullboundarystates}
\varrho_{A_k} &=p \varrho^{\otimes N}_{+,k} +(1-p)\varrho^{\otimes N}_{-,k} ,    \\
\varrho_{B_k} &=\varrho^{\otimes N_+}_{+,k} \otimes \varrho^{\otimes (N-N_+)}_{-,k},
\end{aligned}
\end{equation}
where $\varrho_{\pm,k}:=\tfrac{\id} d \pm \tfrac{1}{2} c_k g_k + \tfrac{1}{2} \sum_{r\neq k} (\aver{G_r}/N) g_r$, with $c_k=\sqrt{\Lambda_{\max}-\sum_{r\neq k} \aver{G_r}^2/N^2}$ for a certain direction $k$ and certain numbers $N_+\in \mathbb N$ and $p\in \mathbb R$, which later will be related by
\be
p = N_+ / N 
\ee
whenever possible. (When this will not be possible there will be a small difference and $p= N_+/N + \delta$ with $\delta \ll 1$).

First of all, we have that, by construction, $\sum_{r=1}^{d^2-1}\trace(g_r \varrho_{\pm,k})^2=\Lambda_{\max}$ holds, and thus the $\varrho_{\pm,k}$ satisfy Eq.~(\ref{localunc}). Therefore, the above are pseudo-separable according to \cref{def:1}.
Furthermore, the matrices $\mathfrak U$ associated to those states are diagonal. The values of the diagonal entries are
\begin{subequations}
\begin{align}
(\mathfrak{U}_{rr})_{A_k}&=N^24p(1-p)\aver{g_r}_{\pm,k}^2 = N^2 (x_r)_{A_k},\\
(\mathfrak{U}_{rr})_{B_k}&=-4N_+(N-N_+) \aver{g_r}_{\pm,k}^2 \tfrac{1}{N-1} = N^2 (x_r)_{B_k}, \label{eq:XrrvarrhoXk} 
\end{align}
\end{subequations}
for $1\leq k \leq d^2-1,$ where $\aver{g_r}_{\pm,k}:=\trace(g_r \varrho_{\pm,k})$. Furthermore, we have that $\aver{g_r}_{\pm,k}=\aver{G_r}/N$ for all $r\neq k$, while we have
\be
\aver{g_k}_{\pm,k}=c_k=\sqrt{\LambdaSigma+\aver{G_k}^2/N^2} \quad \text{with} \quad \LambdaSigma=\Lambda_{\max} -\sum_{r=1}^{d^2-1} \aver{G_r}^2/N^2 .
\ee
Let us also recall the definition of the last coordinate given in \cref{xd2}, 
and let us calculate it for those states. For the states $\varrho_{B_k}$ we have
\be
\begin{aligned}\label{eq:xd2Xk}
	(x_{d^2})_{B_k} &= -4\tfrac{N_+(N-N_+)}{N^2} \tfrac{1}{N-1}\Lambda_{\max} = \tfrac 1 {N^2} \Tr(\mathfrak{U}_{B_k}) ,
\end{aligned}
\ee
where the last equality follows by calculating the sum of the values $(\mathfrak U_{rr})_{B_k}$ in \cref{eq:XrrvarrhoXk}. 

For the states $\varrho_{A_k}$ instead, we find
\be\label{eq:trCYk}
 \tfrac 1 {N} \tr(C_\varrho) =  \left(N +d \right) \Lambda_{\max}
 \quad \Rightarrow \quad  (x_{d^2})_{A_k} = 0 ,
\ee
which is consistent with the fact that the state is permutationally symmetric (bosonic).

By straightforward algebra we also find the following relation
\be
\begin{aligned}\label{eq:trXYk}
\tfrac 1 {N^2}\Tr(\mathfrak{U}_{A_k})&=4p(1-p)\Lambda_{\max} .
\end{aligned}
\ee
Let us now verify the relation \eqref{normconst} for the states $\varrho_{B_k}$. For the right-hand side we have
\be
\begin{aligned}
	N (x_{d^2})_{B_k} + \sum_{k=1}^{d^2-1} (x_{k})_{B_k} &= N (x_{d^2})_{B_k} + \tfrac 1 {N^2}  \Tr(\mathfrak{U}_{B_k}) 	
	= \frac{N_+(N-N_+)}{N^2} \Lambda_{\max},
\end{aligned}
\ee
which follows from \cref{eq:xd2Xk}.
The left-hand side instead, is simply the definition of $\LambdaSigma$ that is
\be
\LambdaSigma = \Lambda_{\max} - |\aver{\vec G}|^2/N^2.
\ee
Thus, in order to verify \cref{normconst} for the states $\varrho_{B_k}$ we have to impose
\be
4N_+(N-N_+)/N^2 = \LambdaSigma/\Lambda_{\max} := \kappa ,
\ee
where in the latter equality we defined the quantity $\kappa,$ for which $0\le\kappa\le1$ holds.

Similarly, for the states $\varrho_{A_k}$ we have that the right-hand side of \cref{normconst} is given by
\be
N (x_{d^2})_{A_k} + \tfrac 1 {N^2}  \Tr(\mathfrak{U}_{A_k}) = 4p(1-p)\Lambda_{\max} ,
\ee
due to \cref{eq:trCYk,eq:trXYk}.
Thus, in this case in order to verify \cref{normconst} we have to impose
\be
4p(1-p) = \LambdaSigma/\Lambda_{\max}= \kappa .
\ee

In summary, to have a correspondence $(\varrho_{A_k},\varrho_{B_k})\rightarrow (A_k,B_k)$ for a fixed value of $\aver{\vec G}$ we have to impose the following condition
\begin{align}
4N_+(N-N_+)/N^2 = 4p(1-p) &= \LambdaSigma/\Lambda_{\max} := \kappa ,
\end{align} 
that can be satisfied {\it if and only if} $Np=N_+$ is an {\it integer number}. In particular its value will be related to $\LambdaSigma$ as 
\be
N_+=\frac N 2 \left( 1 \pm \sqrt{1-\LambdaSigma/\Lambda_{\max}} \right). 
\ee
When $Np$ is not integer we consider
$(\varrho_{A_k},\varrho_{B^\prime_k})$ such that $Np=N^\prime_+$ and
\begin{equation}\label{condonsurf1}
N^\prime_+ =\bigg\lfloor \frac N 2 \left(1 \pm \sqrt{1-\LambdaSigma/\Lambda_{\max}  } \right)\bigg\rfloor := N_+ - \epsilon,
\end{equation}
where $\lfloor x \rfloor$ is the integer part of $x$ and we defined 
\be
\epsilon := N_+ - N^\prime_+  \quad \text{with} \quad 0\leq \epsilon \leq 1 .
\ee
Then again $\varrho_{A_k}$ will correspond to $A_k$ and
$\varrho_{B_k^\prime}$ will correspond to a point $B_k^\prime$ with coordinates 
\be
\begin{aligned}
(x_{d^2})_{B_k^\prime} &= -4\frac{N^\prime_+(N-N^\prime_+)}{N^2(N-1)}\Lambda_{\max} = (x_{d^2})_{B_k} + 4\frac{\epsilon (N + \epsilon )}{N^2(N-1)}\Lambda_{\max}  = (x_{d^2})_{B_k} + O(1/N^2),  \\
(x_{r})_{B_k^\prime} &= -4\frac{N^\prime_+(N-N^\prime_+)}{N^2(N-1)} \aver{g_r}_{\pm,k}^2 = (x_{d^2})_{B_k} + 4\frac{\epsilon (N + \epsilon )}{N^2(N-1)}\aver{g_r}_{\pm,k}^2 = (x_{r})_{B_k} + O(1/N^2),
\end{aligned}
\ee
where $O(\cdot)$ is the usual big-$O$ notation that describes the asymptotic behavior for large $N$ in this case.

From this, we can observe that the signed distance between the two points $B_k$ and $B_k^\prime$, is given by 
\be
\| B_k-B_k^\prime \|=\sum_{i=1}^{d^2} | (x_{r})_{B_k^\prime} - (x_{r})_{B_k} | = O(1/N^2) .
\ee
Now, one might ask whether such a scaling is negligible or not in the large $N$ limit, as compared to the size of the polytope. 
For that, we can consider as a length scale the smallest distance between two points $X_{k}$ and $X_l$ which would correspond to the smallest edge. Note in fact that, for instance the distance between a point $B_k$ and a point $A_k$ is much longer. See \cref{fig:polytope}.
From \cref{extrpoints} we find that
\be
\| B_k-B_l \|=\sum_{i=1}^{d^2} | (x_{r})_{B_k} - (x_{r})_{\rm B_l} | = \frac{\kappa}{N-1} (|c_k^2 - \aver{G_k}^2/N^2 | +  |c_l^2 - \aver{G_l}^2/N^2 |) 
=  2\frac{\kappa}{N-1} \Lambda  ,
\ee
which is of order $O(1/N)$.
Here, we consider this specific distance because it is the one directly related to our entanglement parameter.
Note however, that a similar observation can be made for other distances.

Thus in the limit $N\gg 1$ the two points $B_k$ and $B_k^\prime$ coincide. This means that in the limit
$N\gg 1$ there are pseudo-separable matrices corresponding to all vertices of $\OmegaSigma$ for all $\aver{\vec G}$. \qed 

For completeness, we provide in the following also the average two-body reduced states of the states at the vertices of the $su(d)$-polytope:
\begin{subequations}
\begin{align}
\rhoavtwo^{A_k} &= p \varrho_{+,k}\otimes \varrho_{+,k} +(1-p)\varrho_{-,k} \otimes \varrho_{-,k},  \\ 
\rhoavtwo^{B_k} &= \tfrac{p(N_+-1)}{N-1} \varrho_{+,k}\otimes \varrho_{+,k} +
\tfrac{Np(1-p)}{N-1} (\varrho_{+,k}\otimes \varrho_{-,k} 
+ \varrho_{-,k}\otimes \varrho_{+,k})+ \tfrac{(1-p)(N-N_+-1)}{N-1} \varrho_{-,k}\otimes \varrho_{-,k} .
\end{align}
\end{subequations}
Note once more that $\rhoavtwo^{A_k}$ satisfies $\aver{F}=1$ (corresponding to $x_{d^2}=0$), meaning that they are bosonic states, while $\rhoavtwo^{B_k}$ satisfies $\Tr(\barX)=\tfrac 2 N (1-\aver{F})$ (corresponding to $x_{d^2}=-\sum_{l=1}^{d^2-1} x_l$), which is valid for all product states. 

\section{Details of numerical calculations}\label{app:nuemrics}

We used MATLAB \cite{MATLAB2020} for numerical calculations. We also used the QUBIT4MATLAB package \cite{Toth2008QUBIT4MATLAB,QUBIT4MATLAB_actual_note62_href}.
The pseudo-separable state given in \cref{eq:entPSsep} is obtained with {\tt pseudoseparable\_state3x3.m}. The use of the command is demonstrated by {\tt example\_pseudoseparable.m}.
QUBIT4MATLAB includes the routine {\tt optspinsq.m} computing the optimal-spin squeezing inequalities 
as well as {\tt optsudsq.m} calculating the $su(d)$-squeezing inequalities. The function {\tt optsudsq\_jxjzjz.m} uses the $su(d)$ generators given in \cref{eq:gk}. Moreover, 
{\tt example\_fully\_connected\_sud\_chain.m} can be used to reproduce Table 1, and 
{\tt example\_fully\_connected\_spin\_chain.m} can be used to reproduce Tables 2 and 3.
Finally, {\tt example\_optsudsq\_jxjyjz.m} reproduces 
\cref{fig:Ueig}.

\bibliographystyle{quantum}
\bibliography{Bibliography.bib}

\end{document}